\documentclass[]{aa}

\usepackage{graphicx}
\usepackage{amsmath}
\usepackage{txfonts}
\usepackage{subfigure}
\usepackage{siunitx}
\usepackage{xspace}
\usepackage{natbib}
\usepackage{url}
\usepackage{color}

\newcommand*{\degree}{\si{\degree}}
\newcommand*{\arcminute}{\si{\arcminute}}
\newcommand*{\diff}{\ensuremath{{\rm d}}}

\newcommand*{\Euclid}{\textit{Euclid}\xspace}

\begin{document}

\defcitealias{planck2018}{PC18}

\title{Higher order statistics of shear field: a machine learning approach}
\author{
Carolina Parroni\inst{1}
\and
\'Edouard Tollet\inst{2,3}
\and
Vincenzo F. Cardone\inst{1,4} 
\and
Roberto Maoli\inst{5}
\and
Roberto Scaramella\inst{1,4}
}

\institute{
I.N.A.F. - Osservatorio Astronomico di Roma, via Frascati 33, 00040 - Monte Porzio Catone (Roma), Italy 
\and
Observatoire de Paris, GEPI and LERMA, PSL University, 61 Avenue de l'Observatoire, F-75014 Paris, France; Universit\'e Paris Diderot, Sorbonne Paris Cit\'e, F-75013 Paris, France
\and
Centre de Recherche Astrophysique de Lyon UMR5574, ENS de Lyon, Univ. Lyon1, CNRS, Universit\'e de Lyon, 69007, Lyon, France
\and
Istituto Nazionale di Fisica Nucleare - Sezione di Roma 1, Piazzale Aldo Moro, 00185 - Roma, Italy
\and
Dipartimento di Fisica, Universit\`a di Roma "La Sapienza", Piazzale Aldo Moro, 00185 - Roma, Italy}

\abstract
{The unprecedented amount and the excellent quality of lensing data that the upcoming ground- and space-based surveys will produce represent a great opportunity to shed light on the questions that still remain unanswered concerning our universe and the validity of the standard $\Lambda$CDM cosmological model. Therefore, it is important to develop new techniques that can exploit the huge quantity of data that future observations will give us access to in the most effective way possible.}
{For this reason, we decided to investigate the development of a new method to treat weak lensing higher order statistics, which are known to break degeneracy among cosmological parameters thanks to their capability of probing the non-Gaussian properties of the shear field. In particular, the proposed method directly applies to the observed quantity, i.e., the noisy galaxy ellipticity.}
{We produced simulated lensing maps with different sets of cosmological parameters and used them to measure higher order moments, Minkowski functionals, Betti numbers, and other statistics related to graph theory. This allowed us to construct datasets with different size, precision, and smoothing. We then applied several machine learning algorithms to determine which method best predicts the actual cosmological parameters associated with each simulation.}
{The best model resulted to be simple multidimensional linear regression. We used this model to compare the results coming from the different datasets and found out that we can measure with good accuracy the majority of the parameters that we considered. We also investigated the relation between each higher order estimator and the different cosmological parameters for several signal-to-noise thresholds and redshifts bins. }
{Given the promising results, we consider this approach as a valuable resource, worth of further development.}


\maketitle

\section{Introduction}

During the past decades, the availability of multi-band astronomical data of ever increasing quality has lead to an impressive progress in the field of observational cosmology that resulted in the establishment of the concordance $\Lambda$CDM model. Its parameters (specifying the contribution of matter and cosmological constant to the energy budget, its expansion rate, and growth of structures) have been measured with an unprecedented level of precision through the joint use of different cosmological probes such as the angular anisotropy of the cosmic microwave background (CMB), the baryon acoustic oscillation (BAO), galaxy clustering (GC), and weak lensing (WL) \citetext{e.g. \citealp{dunkley2009}, Planck Collaboration \citeyear{planck2014, planck2016, planck2018}, \citealp{alam2017}, Dark Energy Survey (DES) Collaboration \citeyear{abbott2018}}.

In particular, Planck Collaboration \citeyearpar{planck2018} (hereafter PC18) combined measurements of CMB polarization, temperature, and lensing, with BAO and type Ia supernovae (SNe) data in order to obtain the tightest possible constraints on the cosmological parameters. \citetalias{planck2018} constraints are in good agreement with different BAO, SNe, and some galaxy lensing observations, but they show a slight tension with the Dark Energy Survey (DES) Collaboration \citeyearpar{abbott2018} results, obtained including GC and WL data, and they also present a more significant tension with the local measurements of the Hubble constant \citep{riess2018}. Considering the precision of these measurements and the accuracy with which those studies were performed, one could speculate that the observed tensions may be linked to new physics, or to phenomena not accounted for in the standard cosmological model, more than to systematic errors. In fact, despite the successful results that were obtained, it is important to stress that the nature of the main energy contents of the universe predicted by the $\Lambda$CDM model (i.e. dark energy that drives cosmic speed up and dark matter, responsible for the formation of large-scale structures) still remains unknown.
  
Ongoing surveys, e.g. DES \citetext{DES Collaboration \citeyear{des2005}}, the Hyper Suprime-Cam \citep[HSC,][]{aihara2017}, and the Kilo-Degree Survey \citep[KiDS,][]{dejong2012}, and next-generations surveys from the ground, such as the Legacy Survey of Space and Time \citep[LSST,][]{lsst2009}, or space-based, like the ESA \Euclid \citep{laureijs2011} and the NASA Wide Field Infrared Survey Telescope \citep[WFIRST,][]{green2012} missions have the goal of shedding light on the many questions that still remain open. Notably the \Euclid survey will collect both imaging and spectroscopic datasets using WL and galaxy clustering as primary probes in order to constrain with unprecedented precision the dark energy equation of state, measure the rate of cosmic structure growth to discriminate between General Relativity against and Modified Gravity, and look for deviations from Gaussianity of initial density perturbations to test inflationary scenarios. In particular, \Euclid will obtain high-quality data on sub-arcsec scale of galaxy shape measurements for galaxies up to $z\geq2$, covering $\num{15000}\,\rm{deg}^2$ of the extragalactic sky.  
   
Although on large scales the density field is well approximated by a Gaussian distribution, the information brought by the measurement of non-Gaussianity on small scales can help to break degeneracies and further constrain the cosmological parameters. WL is considered as one the best tools for accessing this information. As predicted by General Relativity (and any metric theory of gravity), the matter distribution along the line-of-sight deflects the light rays because they propagate along the geodesic lines, causing a distortion of the image of the emitting sources. In the WL regime, this effect is too small to be detected on single galaxies, and a statistical approach is needed to access the information contained in the cosmic shear field.

Given its sensitivity to the background expansion and to the growth of structures, lensing second-order statistics have been employed with remarkable success in the past, through the analysis of the two-point correlation function, and its Fourier counterpart, the power spectrum \citep[see, e.g.,][and references therein]{munshi2008, kilbinger2015, bartelmann2017, kohlinger2017, hildebrandt2017, troxel2018, hikage2019, hamana2020}. In order to access the non-Gaussian information originating from the nonlinear collapse of the primordial density fluctuations though, it is necessary to go higher order in the statistical description of the shear field. Along with the more traditional three- and four-point correlation functions, and the corresponding bi- and tri-spectra in Fourier space, \citep[e.g.][]{takada2003, takada2004, semboloni2011, fu2014} various estimators have been more recently used. Topological descriptors such as Minkowski functionals and Betti numbers have been applied to lensing convergence maps \citep{matsubara2001, sato2001, taruya2002, matsubara2010, kratochvil2011, pratten2012, petri2013, shirasaki2014, ling2015, vicinanza2019, marques2019, mawdsley2020, parroni2020, zurcher2020}, and on three-dimensional Gaussian random fields to study the topology of the primordial density field \citep{park2013, pranav2017, pranav2019}, respectively. Moreover \citet{hong2020} applied graph theory estimators to study the topological structure of clustering in N-body simulations corresponding to different cosmological models.   
  
In the context of WL studies, the topological higher order estimators have usually been applied to lensing convergence maps. However, the convergence is not a direct observable so that one has actually to solve an inversion problem starting from the shear data. Even if several methods have been conceived and a lot of progress has been made \citep[e.g.][]{pires2009,jullo2014,jeffrey2018,pires2019, price2020a, price2020b}, this reconstruction is still considered as a non-trivial problem that requires very accurate control of systematic effects coming from survey masking, borders, noise, and the fact that what we actually observe is the galaxy ellipticity, which is a measure of the noisy \textit{reduced} shear and not the shear itself. In order to circumvent this reconstruction problem, higher order statistics could be applied directly on ellipticity maps but the issue in this case would be the lack of theoretical predictions to compare the measurements with. Even in the cases where a theoretical study has indeed been carried out, as for the higher order moments and Minkowski functionals of convergence maps, we have to take into account the approximated calculations due to the challenge of modeling non-linearities in the matter power spectrum and bispectrum. The mismatch between theoretical expected values and actual noisy observations can be dealt with through a calibration process performed on simulations \citep[e.g.][]{vicinanza2018,vicinanza2019,parroni2020} but this requires to create an appropriate parametrization, which is nevertheless a non trivial approximation adding nuisance parameters and possible degeneracy with the cosmological ones hence weakening the constraints.
  
For these reasons we decided to apply higher order moments, Minkowski functionals, Betti numbers, and several statistics from graph theory on simulated noisy ellipticity maps, and to use different machine learning techniques to study the relation between those estimators and the cosmological parameters that were used to generate the simulations. In recent years, machine learning has proven to be a valuable tool in a variety of astrophysical studies and notably in some WL applications such as the discrimination between different modified gravity cosmologies \citep[e.g. ][]{merten2019, peel2019}, the measurement of the ($\Omega_{\rm{M}}$,$\sigma_8$) parameters degeneracy \citep[e.g. ][]{gupta2018, fluri2018, fluri2019}, and mass maps reconstruction \citep[e.g. ][]{jeffrey2020}. In our case, it allows us to bypass the theory issues that we discussed, and it lets us make direct use of noisy ellipticity maps, on which we calculated new and promising higher order estimators.    
  
The paper is organized as follows: in Section\,\ref{sec:simulations} we describe how we obtained the simulated shear maps for the different set of cosmological parameters. In Section\,\ref{sec:estimators} we introduce the different higher order estimators that we measured on the maps and we present the final datasets that we used for the training phase. In Section\,\ref{sec:model_sel} we compare the results from different models, in Section\,\ref{sec:prediction} we use the best model obtained to study the effect of the dataset size, of the measurements accuracy, and of the smoothing scale on the score, performing the training and the predictions using the different datasets, and in Section\,\ref{sec:feat_imp} we study the relation between the individual estimators and each cosmological parameter. In Section\,\ref{sec:discussion} we discuss the limitations of this work and the possible improvements. In Section\,\ref{sec:conclusions} we draw our conclusions. In Appendix \,\ref{app:models} we briefly outline the different machine learning methods that we compared. 

\section{Simulation of shear maps}\label{sec:simulations}

In order to produce the simulated shear maps, we used {\tt The Full-sky Lognormal Astro-fields Simulation Kit} \citep[\texttt{FLASK};][]{flask}. \texttt{FLASK} is a fast and flexible public code that, taking as input the auto- and cross-power spectra, is able to create random realizations of different astrophysical fields that follow a multivariate lognormal distribution, reproducing the expected cross-correlations between the input fields.

The choice of a multivariate lognormal distribution is motivated by the better approximation that this distribution represents of the fields that we want to simulate, compared to a multivariate Gaussian distribution \citep[e.g. ][]{scaramella1993, taruya2002, hilbert2011, clerkin2017}. This is also the simpler approximation that can convey the non-Gaussian information contained in the shear field, that we are interested to measure. Moreover, a non-negligible aspect of this kind of simulation is the computational speed that it offers, which allows \texttt{FLASK} to produce full-sky realization within minutes.

While we refer to \citet{flask} for the details on \texttt{FLASK} inner workings, we want to stress here some limitations of this approach. In one of the two proposed solutions, \texttt{FLASK} computes the shear starting from the convergence, which is in turn calculated through an approximated line-of-sight integration of the simulated density field. This affected the choice of the redshift range of the simulations. In fact, this approximation consists in a weighted Riemann sum over the chosen redshift bins, which is able to reproduce the theoretical spectra within $3\%$ for z>0.5. Due to the small number of bins in the sum at low redshift, the precision of the approximation quickly degrades. For this reason, we decided to cut the source catalog at $z>0.55$. Moreover, the line-of-sight integration solution produces a convergence field that follows a distribution of a sum of correlated lognormals, which is not exactly lognormal, even if very similar. Although it is possible to add a shift the convergence field generated by \texttt{FLASK} in order to match the third order moment, this would artificially alter the convergence pdf so that moments higher than the third would be modified in an unpredictable way. Moreover, lacking a theoretical estimate for some of the statistics we will consider below, there is actually no way to judge whether \texttt{FLASK} is able to reproduce them. However, running a number of full N - body and ray tracing simulations as large as the one we need for our study is definitely not possible with the computing resources at our disposal. We therefore prefer to rely on \texttt{FLASK} for this preliminary study since we want to show the potentiality of the method we are proposing rather than applying it to real data.

We chose a \Euclid-like source redshift distribution as in Euclid Collaboration \citeyearpar{IST2019}

\begin{equation}
n(z) = \frac{3 \, n_{\rm{g}}}{2 \, z_0} \, \left ( \frac{z}{z_0} \right )^2 \,\exp{ \left [ -\left ( \frac{z}{z_0} \right )^{3/2} \right ]}
\label{eq:nz}
,\end{equation}
with $n_{\rm{g}}$ the number of galaxies per $\rm{arcmin}^{2}$, and $z_{0}  = z_{\rm{m}}/\sqrt{2}$ with $z_{\rm{m}}$ the median redshift. We set $n_{\rm{g}}=30\,\rm{gal/{arcmin}^{2}}$ and $z_{\rm{m}}=0.9$ as expected for \Euclid given a limiting magnitude $\rm{mag_{lim}} = 24.5$ in the imaging VIS filter.

\begin{table}
\begin{center}
\begin{tabular}{lll}
\hline \hline
parameter & $\mu$ & $\sigma$ \\
\hline
$\rm{H_0}$ & 67.66 & 0.42 \\
$\omega_{\rm{b}}$ & 0.02242 & 0.00014 \\
$\Omega_{\rm{M}}$ & 0.3111 & 0.0056 \\
$\Omega_\Lambda$ & 1-$\Omega_{\rm{M}}$ & 0.0056 \\
$\rm{w_0}$ & -1.028 & 0.032 \\
$\rm{n_s}$ & 0.9665 & 0.0038 \\
$\sigma_8$ & 0.8102 & 0.0060 \\
\hline\hline
\end{tabular}
\end{center}
\caption{Mean and width of the Gaussian distributions from which we extracted the cosmological parameter values for each simulation, corresponding to \citetalias{planck2018} results and errors.
}
\label{tab:planck}
\end{table}

We used \texttt{CLASS} \citep{blas2011,dio2013} to compute the input power spectra for 25 top-hat equi-spaced redshift bins over the range $0.0 \le z \le 2.5$ for a flat $\Lambda$CDM model, varying the cosmological parameters $\{ \rm{H_0}, \omega_{\rm{b}}, \Omega_{\rm{M}}, \Omega_\Lambda, \rm{w_0}, \rm{n_s}, \sigma_8 \}$, in each simulation. Because we wanted to compare our measurements and errors on each of those parameters with state of the art results, we chose to refer to the results presented in \citetalias{planck2018}. For each simulation, we randomly extracted each parameter from a Gaussian distribution with mean $\mu$ and width $\sigma$, corresponding to the measured values and errors of \citetalias{planck2018} parameters, respectively, which are shown in Table\,\ref{tab:planck}. With these settings, we created a first batch of $\sim 1000$ simulations, followed by a second batch of $\sim 500$ simulations for which we doubled the width of each Gaussian, in order to cover a larger area in the parameter space. 

Setting $\rm{N_{side}}=2048$ and giving as input to \texttt{FLASK} the source redshift distribution, the mask based on the private \Euclid Flagship galaxy mock catalog version 1.6.18, and the angular auto- and cross-power spectra calculated as described, we obtained a catalog containing the coordinates, the redshift, and the noisy ellipticity components for each galaxy, for each simulation.

As described in \citet{flask}, the complex ellipticity $\varepsilon = \varepsilon_{1} + i\varepsilon_{2}$ is computed as
\begin{equation}
\varepsilon= \left \{ 
\begin{array}{ll}
\displaystyle{\frac{\varepsilon_{s}+g}{1+g^{*}\varepsilon_{s}}}, & \displaystyle{|g|\leq1} \\
& \\
\displaystyle{\frac{1+g\varepsilon_{s}^{*}}{\varepsilon_{s}^{*}+g^{*}}}, & \displaystyle{|g|> 1} \\
\end{array}
\right .
\label{eq:ellipticity}
\end{equation}
where $g\equiv\gamma/(1-\kappa)$, is the reduced shear, $\gamma$ is the shear, and $\kappa$ is the convergence, and $\varepsilon_{s}=\varepsilon_{s,1}+i\varepsilon_{s,2}$ is the source intrinsic ellipticity, whose components $(\varepsilon_{s,1},\varepsilon_{s,2})$ are randomly drawn from a Gaussian distribution with zero mean and a standard deviation $\sigma_{\varepsilon_{s}}$ that can be set by the user. We set $\sigma_{\varepsilon_{s}}=0.3$\footnote{{\it While in the literature, noise is usually added to $\kappa$ or $\gamma$ directly on the map pixels assuming a Gaussian distribution with zero mean and width $\sigma_n=\sigma_{\varepsilon_{s}}/\sqrt{n_g A_{pixel}}$, with $\sigma_{\varepsilon_{s}}=0.3$, FLASK assigns the noise, in the form of $\varepsilon_{s}$, on each individual galaxy trough Eq.\;\ref{eq:ellipticity}.}}. When a map is constructed, the ellipticity components are averaged inside each pixel in order to decrease the noise given by the intrinsic ellipticity. We clarify then that what we call shear maps are therefore maps of the noisy ellipticity, which is the only WL direct observable. More precisely, the value in each pixel is $|\varepsilon| = \varepsilon_{1}^2 + \varepsilon_{2}^2$, the mean of the norm of the complex ellipticity.

Each catalog was then split into redshift bins with equal width $\Delta z = 0.05$ and centered in $z$ from 0.5 to 1.8 in steps of 0.3. According to the \Euclid Red Book \citep{laureijs2011}, photometric redshifts will be measured with an error $<0.05(1+z)$ allowing to separate sources in bins with a center determined with an accuracy better than $0.002(1+z)$ as confirmed by more recent analyses (Euclid Collaboration \citeyear{IST2019}; \citealt{joshi2019}; Euclid Collaboration \citeyear{euclid2020}). As such, our choice $\Delta z = 0.05$ is well within the realistic capabilities of \Euclid. For each slice of redshift, we first obtained $100$ maps, $5 \times 5 \, \rm{deg}^2$ and $300 \times\,300$ pixels leaving a gap of $\sim 1 \, \degree$ between two consecutive maps so that we were able to consider them as independent realization. We performed a gnomonic projection to project the maps onto the plane of the sky, under a flat sky approximation, which holds for the size of the maps that we used. 

\begin{figure}
\centering
\includegraphics[scale=.50]{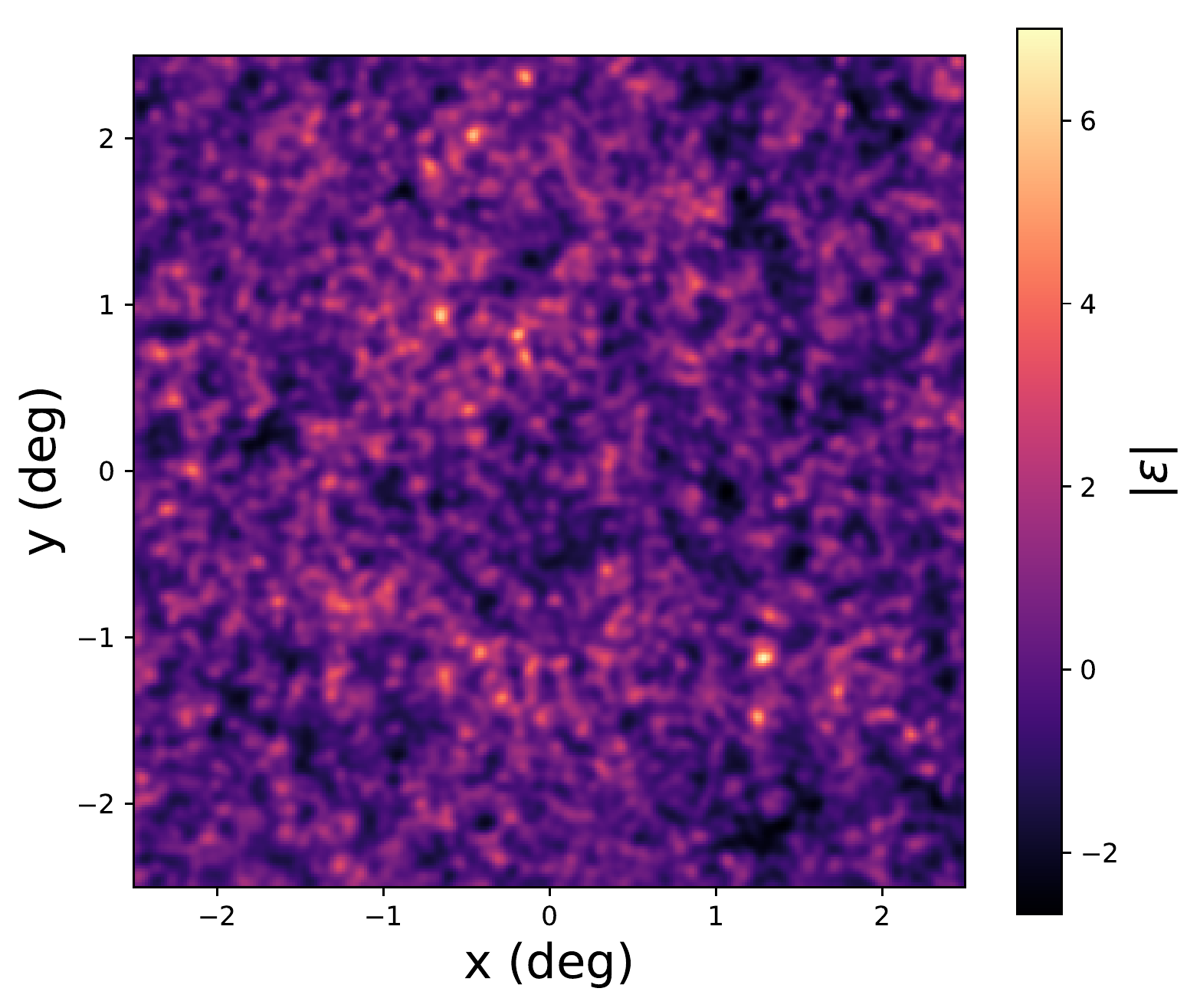}
 \caption{One of the maps taken from the simulation corresponding to \citetalias{planck2018} parameters, at $z=0.6$, smoothed with a Gaussian filter with scale $\theta_{s}=2\arcminute$ and normalized subtracting the mean and dividing by the variance.}
\label{fig:map_planck}
\end{figure}

In Fig.\,\ref{fig:map_planck}, we show one of the shear maps obtained using the catalog from the simulation corresponding to \citetalias{planck2018} parameters, and redshift bin $z=0.6$. The map was smoothed using a Gaussian filter with scale $\theta_{s}=2\arcminute$ and normalized subtracting the mean and dividing by the variance.

In a second moment, in order to increase the training set and the signal-to-noise ratio, we decided to perform four more realizations for each combination of the cosmological parameters already calculated, making therefore $500$ maps for each simulation, bringing the total to $\sim \num{750000}$ maps per redshift bin.

\section{Higher order estimators}\label{sec:estimators}

The simulated maps are the input for our investigation of the potentiality of high order statistics to constrain cosmological parameters. Going beyond second order opens up a wide range of possible choices, and it is not clear {\it a priori} which is the most promising one. For this reason, we considered many different alternatives which we briefly describe in the following paragraphs.

\subsection{Higher order moments}\label{subsec:hom}

Before measuring the higher order moments (HOM), we smoothed the shear maps using a Gaussian filter with scale $\theta_{s}=\{2\arcminute,4\arcminute,6\arcminute\}$ and subtracted the mean value to put all the maps to the same null mean value. Denoting with $\varepsilon(x, \, y)$ the resulting field, on each map, for all redshift bins, we estimated  the third and fourth order centered moments, the skewness, and the kurtosis, respectively defined as:

\begin{equation}
\begin{aligned}
k_{3} & = \left< \varepsilon^{3} \right> \; , \\
k_{4} & = \left< \varepsilon^{4} \right> \; , \\
S_{3} & = k_{3}/{\left< \varepsilon^{2} \right>}^{3/2} \; , \\
S_{4} & = k_{4}/{\left< \varepsilon^{2} \right>}^{2} \; , \\
\end{aligned}
\end{equation}

\begin{figure*}
\centering
\includegraphics[scale=.50]{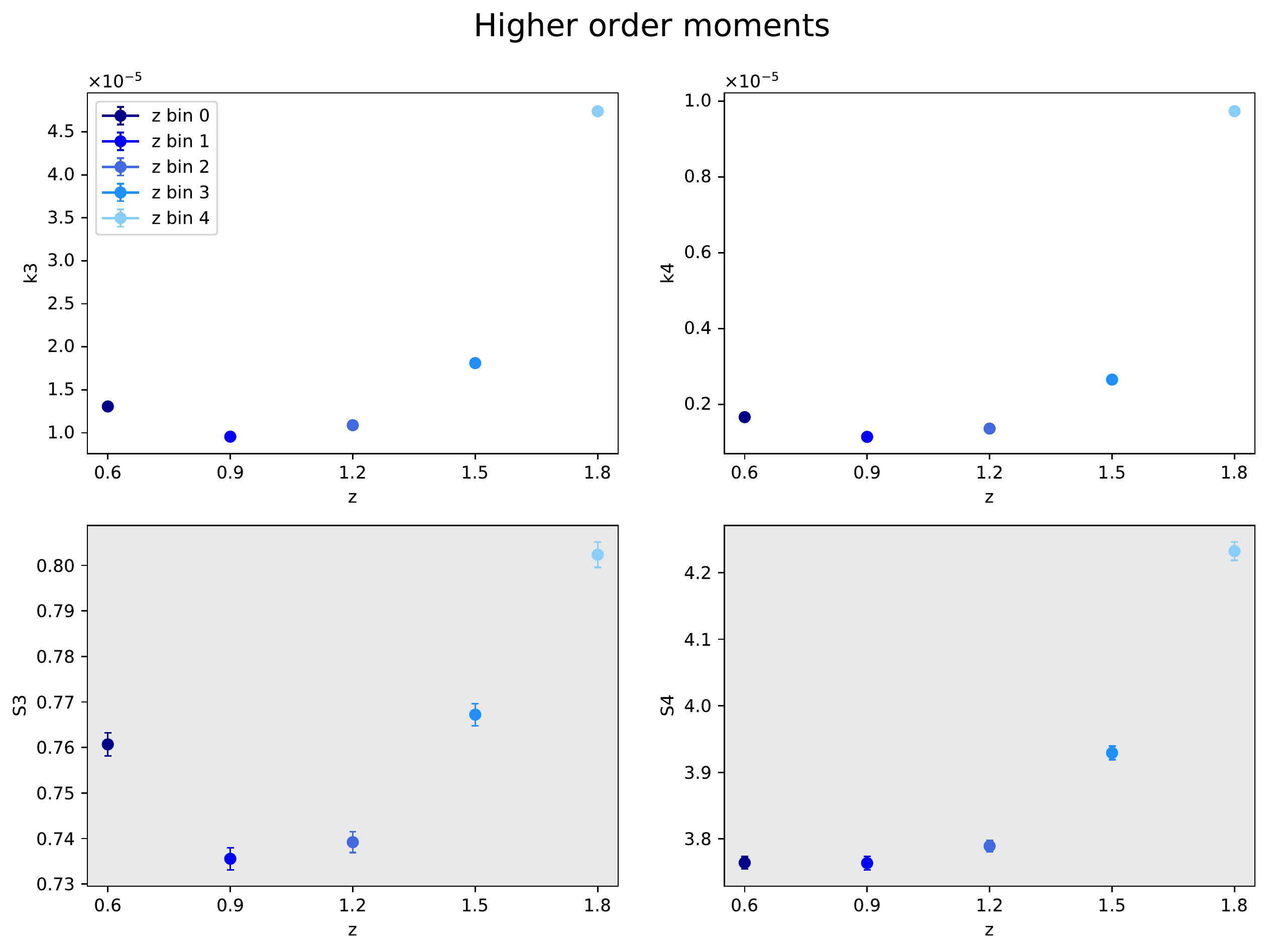}
 \caption{HOM calculated on the maps obtained from the simulation corresponding to \citetalias{planck2018} parameters, averaged over $500$ maps, for all redshift bins, with smoothing scale $\theta_{s}=2\arcminute$. The error bars correspond to the standard deviation divided by the square root of the number of maps. The grayed out plots in the bottom show that the $S_3$ and $S_4$ HOM were discarded from the rest of the analysis.}
\label{fig:hom}
\end{figure*}

In Fig.\,\ref{fig:hom}, we show the HOM as a function of the redshift, calculated on the maps from the simulation corresponding to \citetalias{planck2018} parameters, for the smoothing scale $\theta_{s}=2\arcminute$. The values are averaged over $500$ maps. We notice that all the moments have their minimum value at $z=0.9$, the median redshift of the source distribution, and then they increase with the redshift. We obtained similar results but lower absolute values for the smoothing scales $\theta_{s}=\{4\arcminute, 6\arcminute\}$. In particular, in the simulation corresponding to \citetalias{planck2018} parameters, we measured a decrease in value between the measurements obtained with $\theta_{s}=2\arcminute$ and $\theta_{s}=6\arcminute$ of a factor $\sim 40$. With the exception of $k_4$, for which the minimum value shifts to higher redshift for increasing smoothing, the behavior of the HOM remains unaltered.

The two plots in the bottom, corresponding to $S_3$ and $S_4$, are grayed out to indicate that those HOM were discarded because their scatter among different maps generated with the same cosmology was larger than the difference among maps with different cosmologies. In other words, we did not retain the features for which the ratio between the inter-simulation variance and the intra-simulation variance was smaller than one. In this case, one can not discriminate among cosmological models so that the corresponding probe is not expected to be of any help for our aim. In the following, we used the same criterion to discard several values of the other estimators.

\subsection{Minkowski functionals}\label{subsec:mink}

Given the smoothed two-dimensional shear field $\varepsilon(x, \, y)$ with zero mean and variance $\sigma_0^2$, we define the excursion set $Q_{\nu}$ as the region where $\varepsilon/\sigma_0 > \nu$ holds for a given threshold $\nu$. The three Minkowski functionals (MFs) are defined as

\begin{equation}
\begin{aligned}
V_0(\nu) & = \frac{1}{A} \, \int_{Q_{\nu}}{\diff a} \; , \\
V_1(\nu) & = \frac{1}{4 \, A} \int_{\partial Q_{\nu}}{\diff l} \; , \\
V_2(\nu) & = \frac{1}{2 \, \pi \, A} \int_{\partial Q_{\nu}}{{\diff l \; \cal{K}}} \; , \\
\end{aligned}
\label{eq:MFs}
\end{equation}
where $A$ is the map area, $\partial Q_{\nu}$ the excursion set boundary, $\diff a$ and $\diff l$ are the surface and line element along $\partial Q_{\nu}$, and ${\cal{K}}$ its curvature. Therefore $V_0, \, V_1, \, \text{and }V_2$ are the area, the perimeter, and the genus characteristics of the excursion set $Q_{\nu}$.

Defining $\epsilon\equiv\varepsilon/\sigma_0$, we can rewrite Eq.\,\ref{eq:MFs} as

\begin{equation}
\begin{aligned}
V_{0}(\nu) & = \frac{1}{A} \int_{A}{\diff x \diff y \; \Theta(\epsilon- \nu)} \; , \\
V_{1}(\nu) & = \frac{1}{4 \, A} \int_{A}{\diff x \diff y \; \delta_D(\epsilon - \nu) \, \sqrt{\epsilon_{x}^{2} + \epsilon_{y}^{2}}} \; , \\
V_{2}(\nu) & = \frac{1}{2 \, \pi \, A}
\int_{A}{\diff x \diff y \; \delta_D(\epsilon - \nu) 
\frac{2 \, \epsilon_x \, \epsilon_y \, \epsilon_{xy} - \epsilon_{x}^{2} \, \epsilon_{yy} - \epsilon_{y}^{2} \, \epsilon_{xx}}
{\epsilon_{x}^{2} + \epsilon_{y}^{2}}} \; , \\
\end{aligned}
\end{equation} 

where $\kappa_i = \partial \kappa/\partial x_i$, and $\kappa_{ij} = \partial^2 \kappa/\partial x_i \partial x_j$ with $(i, \, j) = (x, \, y)$, are the first and second derivatives of the field.

\begin{figure}
\centering
\includegraphics[scale=.50]{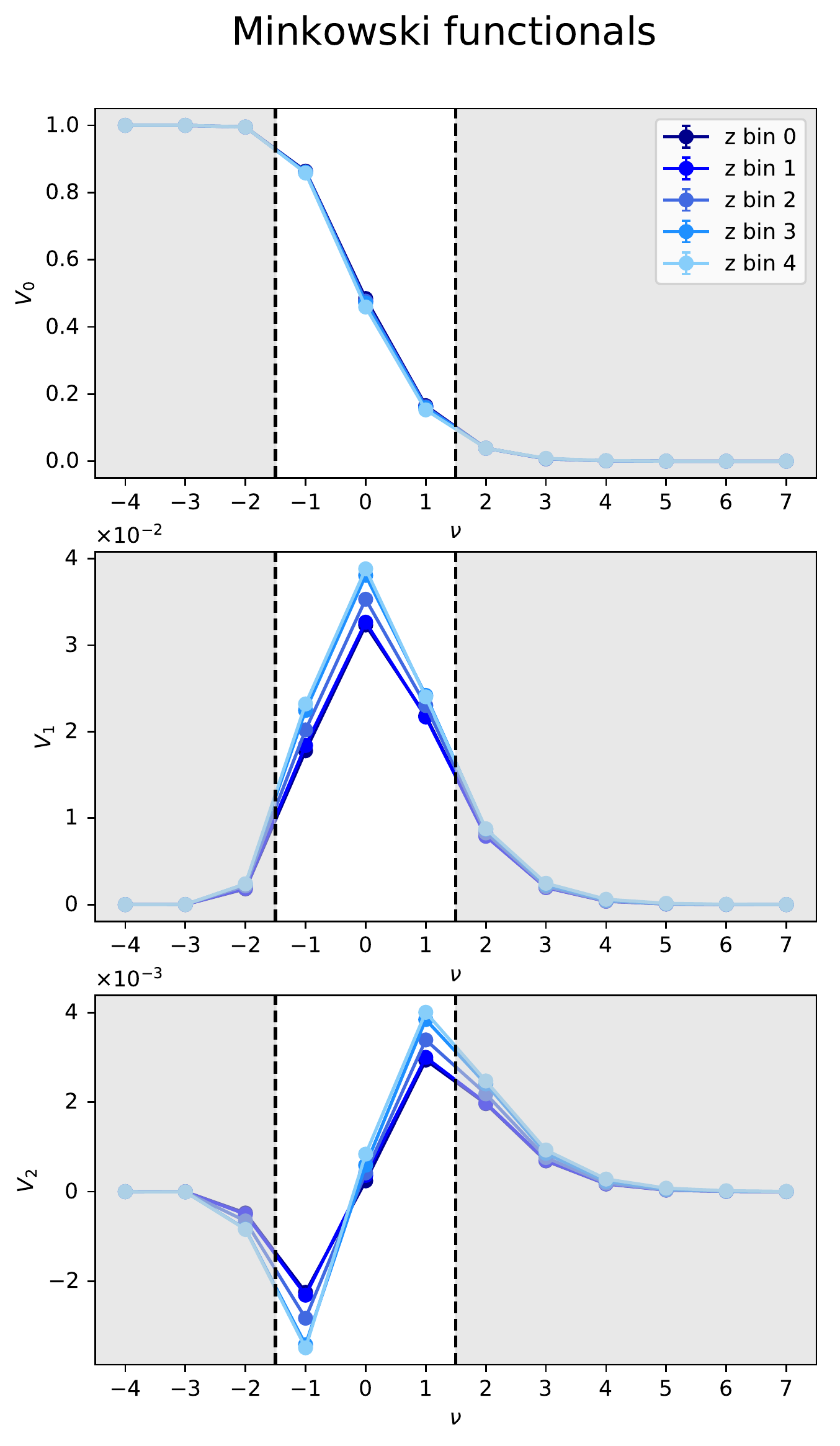}
 \caption{MFs calculated on the maps obtained from the simulation corresponding to \citetalias{planck2018} parameters, averaged over $500$ maps, for all redshift bins, with smoothing scale $\theta_{s}=2\arcminute$. The error bars correspond to the standard deviation divided by the square root of the number of maps. The dashed black lines contain the values selected for the training, the grayed out regions indicate the discarded values.}
\label{fig:mink}
\end{figure}

In Fig.\,\ref{fig:mink}, we can see the three MFs obtained from the maps of the simulation corresponding to \citetalias{planck2018} parameters averaged over $500$ maps, for all redshift bins, with smoothing scale $\theta_{s}=2\arcminute$, as a function of the threshold, with $\nu \in [-4,7]$ and $\Delta \nu = 1$. The dashed black lines delimit the signal-to-noise range that was retained for the training sample, which corresponds to the range $\nu \in [-1,1]$. The values outside of these lines, in the grayed out regions, were discarded following the variance ratio criterion previously defined. While the lines corresponding to $V_0$ at different redshift are almost indistinguishable, we can see that $V_{1}$ values increase with $z$, $V_{2}$ values for $\nu\geq0$ have the same behavior, and $V_{2}$ values for $\nu<0$ tend to decrease as a function of the redshift. For the smoothing scales $\theta_{s}=\{4\arcminute, 6\arcminute\}$, we obtained qualitatively similar results but lower absolute values for $V_{1}$ and $V_{2}$. Specifically, for the simulation corresponding to \citetalias{planck2018} parameters, the measurements of $V_0$ with $\theta_{s}=2\arcminute$ and with $\theta_{s}=6\arcminute$ changed less than $\sim 5\%$, while for $V_{1}$ and $V_{2}$ we observed a decrease in value of a factor $\sim 30$.

\subsection{Betti numbers}\label{subsec:betti}

The definition of Betti numbers \citep{betti1870} requires the knowledge of some fundamental concepts of simplicial homology. A proper treatment of this topic is beyond the scope of this paper so we refer to more specific resources \citep[e.g. ][]{munkres1984,delfinado1993,edelsbrunner2008} for the details and formal definitions.

\begin{figure}
\centering
\includegraphics[scale=.60]{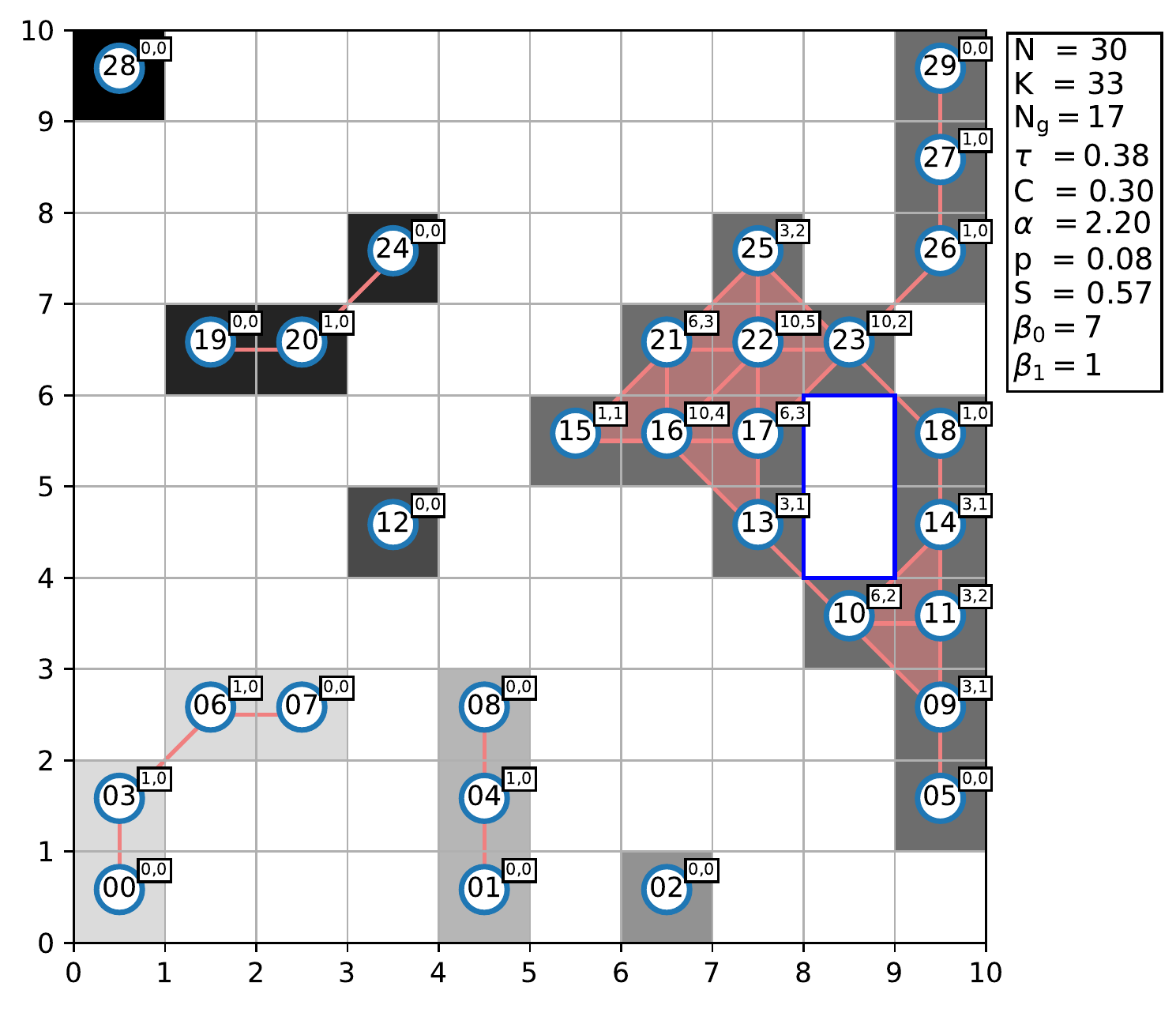}
 \caption{Working example obtained from an excursion set of a random Gaussian field on a $10\times 10$ pixel map, to illustrate the quantities defined in Sec.\,\ref{subsec:betti} for the Betti numbers, and in Sec.\,\ref{subsec:graph} for the graph statistics. Every numbered blue circle is a vertex, the pink lines are edges, the shaded pink regions are triangles. Different connected regions are represented with different shades of gray, and the holes are outlined in dark blue. The tag on each vertex shows the quantities $k_{i}(k_{i}-1)/2$, and $\Delta_{i}$, which are the number of connected triples, and the number of triangles, centered on the vertex. The parameters listed on the right are, in order, $N$ the numbers of vertices, $K$ the number of edges, $N_{g}$ the number of vertices belonging to the giant component, $\tau$ the transitivity, $C$ the LCC, $\alpha$ the average degree, $p$ the edge density, $S$ the fraction of vertices belonging to the giant component, and $\beta_{0}$, $\beta_{1}$ the Betti numbers.}
\label{fig:tri_ex}
\end{figure}

Considering again $\epsilon(x,y)$, and the the excursion set $Q_{\nu}$, we define the Betti numbers in two dimensions, $\beta_{0}$ and $\beta_{1}$, as the number of connected regions and the number of holes in the excursion set, respectively. 

In Fig.\,\ref{fig:tri_ex}, we show a working example obtained from an excursion set of a random Gaussian field on a $10\times 10$ pixel map. We applied a Delaunay triangulation to the map and considered two points as connected if they touch each other horizontally, vertically, or diagonally, i.e. if they are $8$-connected. In Fig.\,\ref{fig:tri_ex}, the vertices are represented as numbered blue circles, the edges as pink lines, and the triangles as shaded pink regions enclosed by three edges. Every connected region is represented with a different shade of gray, and the holes are outlined in dark blue. Because we have seven different connected regions and one hole, the Betti numbers in this case will be $\beta_{0}=7$, and $\beta_{1}=1$. 

\begin{figure*}
\centering
\includegraphics[scale=.50]{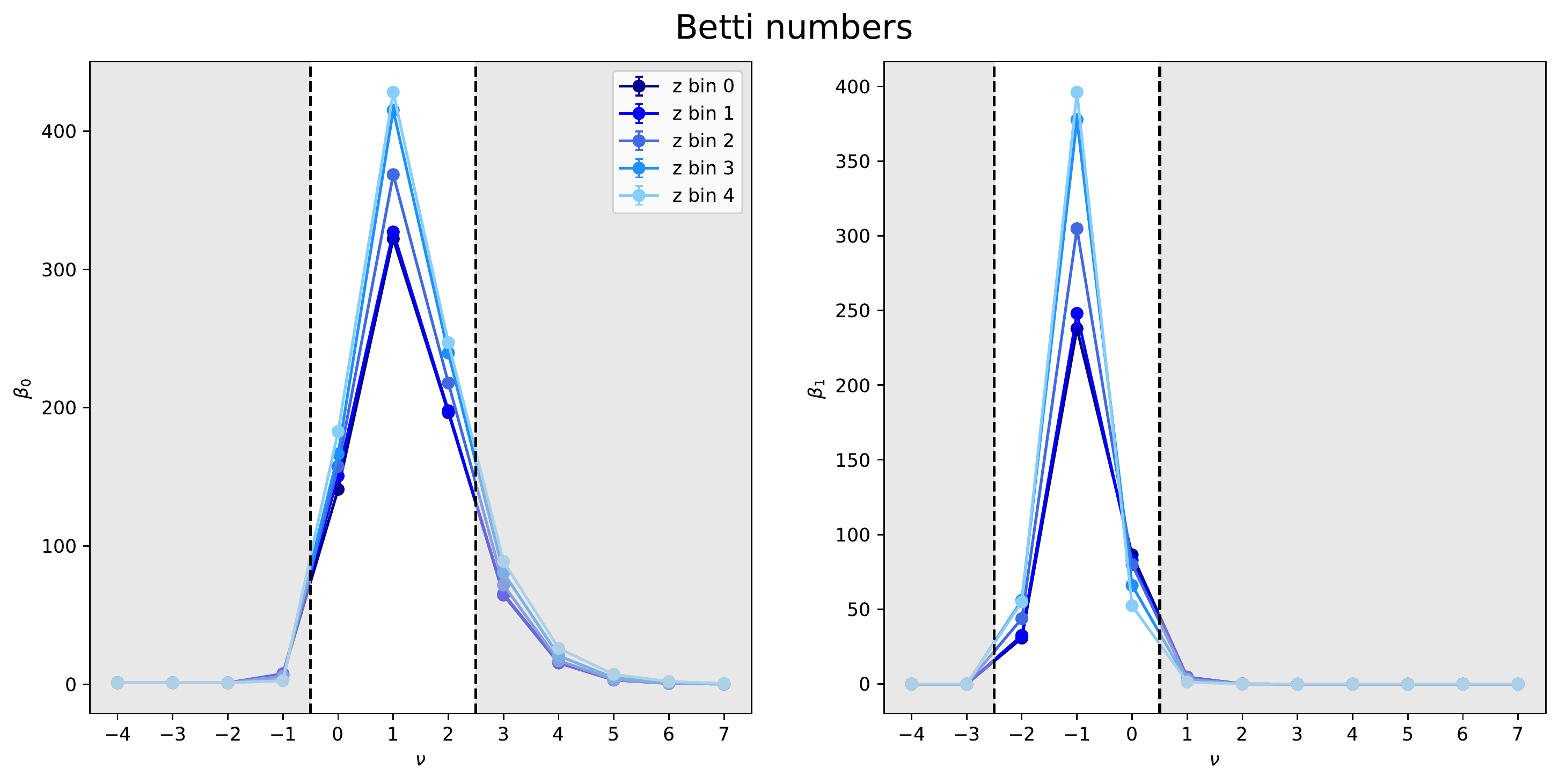}
 \caption{The two Betti numbers calculated on the maps obtained from the simulation corresponding to \citetalias{planck2018} parameters, averaged over $500$ maps, for all redshift bins, with smoothing scale $\theta_{s}=2\arcminute$. The error bars correspond to the standard deviation divided by the square root of the number of maps. The dashed black lines contain the values selected for the training, the grayed out regions indicate the discarded values.}
\label{fig:betti}
\end{figure*}

In Fig.\,\ref{fig:betti}, we show the two Betti numbers as a function of the threshold, with $\nu \in [-4,7]$ and $\Delta \nu = 1$, estimated from the maps of the simulation with \citetalias{planck2018} parameters, averaged over $500$ maps, for the different redshift bins, and with smoothing scale $\theta_{s}=2\arcminute$. As in Fig.\,\ref{fig:mink}, the dashed black lines delimit the threshold range that was retained for the training, i.e. the range $\nu \in [0,2]$ for $\beta_{0}$ and the range $\nu \in [-2,0]$ for $\beta_{1}$, and the grayed out regions indicate the range discarded following the variance ratio criterion. We can see from the left panel, that the number of connected regions for a given threshold increases with the redshift, while in the right panel we note that the number of holes increases with $z$ for $\nu<0$ and it has the opposite behavior for $\nu\geq0$. Increasing the smoothing to scales $\theta_{s}=\{4\arcminute, 6\arcminute\}$, does not change the behavior of the measured curves but their absolute value gets smaller. For the simulation corresponding to \citetalias{planck2018} parameters, the decrease in value between the measurements obtained with $\theta_{s}=2\arcminute$ and with $\theta_{s}=6\arcminute$ is of a factor $\sim 10$.

\subsection{Graph statistics}\label{subsec:graph}

The simplicial complex structure defined in Section \ref{subsec:betti} can also be interpreted as a network or a graph so that some tools used in network science can be applied to it. Following \citet{hong2020} we define the following basic graph quantities

\begin{equation}
\begin{aligned}
\alpha & = \frac{2K}{N} \; , \\
p & = \frac{2K}{N(N-1)} \; , \\
S & = \frac{N_{g}}{N} \; , \\
\end{aligned}
\label{eq:aps}
\end{equation} 

where $N$ is the total number of vertices, $N_{g}$ is the number of vertices belonging to the largest connected sub-graph in a network, called the \textit{giant component}, and $K$ is the total number of edges. Defining the \textit{degree} as the number of neighbors for each vertex, we can call $\alpha$ the \textit{average degree}, while $p$ is the fraction of connected edges over all pairwise combinations and it is therefore called the \textit{edge density}, and $S$ is the fraction of vertices belonging to the giant component. In Fig.\,\ref{fig:tri_ex}, for example, $N=30$ and the largest component is the structure on the right, composed by $17$ vertices, so that $N_{g}=17$ and $S=0.57$. The total number of edges is $K=33$, so that we can calculate also the average degree and the edge density, $\alpha=2.2$ and $p=0.08$.

Given three vertices, if they are connected by at least two edges, they are called a \textit{connected triple}, while if they are connected by three edges, forming therefore a triangle, they are called a \textit{closed triple}. With this definition, a closed triple is also a connected triple. We can look at Fig.\,\ref{fig:tri_ex} to better understand this concept. For example, the vertices $\{19, 20, 24\}$ form one connected triple, centered on the vertex $20$. The vertices $\{00, 03, 06, 07\}$ form two connected triples, one centered on the vertex $03$ and one centered on the vertex $06$. Taking as example the triangle formed by the vertices $\{09, 10 ,11\}$ and ignoring for the moment the other vertices connected to it, i.e. the vertices $05$, $13$ and $14$, we can count three triples which are closed and therefore also connected, $\{11, 09, 10\}$ centered on the vertex $09$, $\{09, 10, 11\}$ centered on the vertex $10$, and $\{10, 11, 09\}$ centered on the vertex $11$. This means that a triangle always counts for three closed triples. Now we can then define quantity

\begin{equation}
\tau = \frac{\text{no. of closed triples}}{\text{no. of connected triples}} = \frac{3 \times \text{no. triangles}}{\text{no. of connected triples}}
\label{eq:tau}
\end{equation} 

which is called \textit{transitivity} or \textit{global clustering coefficient}. In Fig.\,\ref{fig:tri_ex}, $\tau=0.38$. We can also calculate the transitivity for each vertex $i$, a quantity referred to as \textit{local clustering coefficient} (LCC), as

\begin{equation}
  C_{i}=\frac{2\Delta_{i}}{k_{i}(k_{i}-1)}= \frac{\text{no. of closed triples centered on i}}{\text{no. of connected triples centered on i}}
\end{equation}

where $k_{i}$ is the number of neighbors of the vertex $i$, so that $k_{i}(k_{i}-1)/2$ is the total number of connected triples centered on the vertex, and $\Delta_{i}$ is the number of triangles centered on the vertex, i.e. the number of closed triples centered on the vertex. In Fig.\,\ref{fig:tri_ex}, every vertex has a tag with two number, the first corresponds to $k_{i}(k_{i}-1)/2$, and the second is $\Delta_{i}$. For example, the vertex $04$ has two neighbors so that $k_{04}=2$ and $k_{04}(k_{04}-1)/2=1$, and it has no triangles centered on it so that $\Delta_{04}=0$ and therefore $C_{04}=0$. The vertex $16$ has $k_{16}=5$, $k_{16}(k_{16}-1)/2=10$, $\Delta_{16}=4$, so that $C_{16}=0.4$. For vertices like $02$, $12$, and $28$, which have no neighbors and no triangles centered on them, $C_{i}$ is not defined. 

We define the \textit{average} LCC as the averaged value of $C_{i}$ over all $N$ vertices, and call it $C$. In Fig.\,\ref{fig:tri_ex}, $C=0.3$.

We measured $\tau$, $C$, $\alpha$, $p$, and $S$ on the graph obtained by applying a Delaunay triangulation to the $\epsilon(x,y)$ maps that we first downgraded to $100 \times 100$ pixels for computational speed reasons.

\begin{figure}
\centering
\includegraphics[scale=.50]{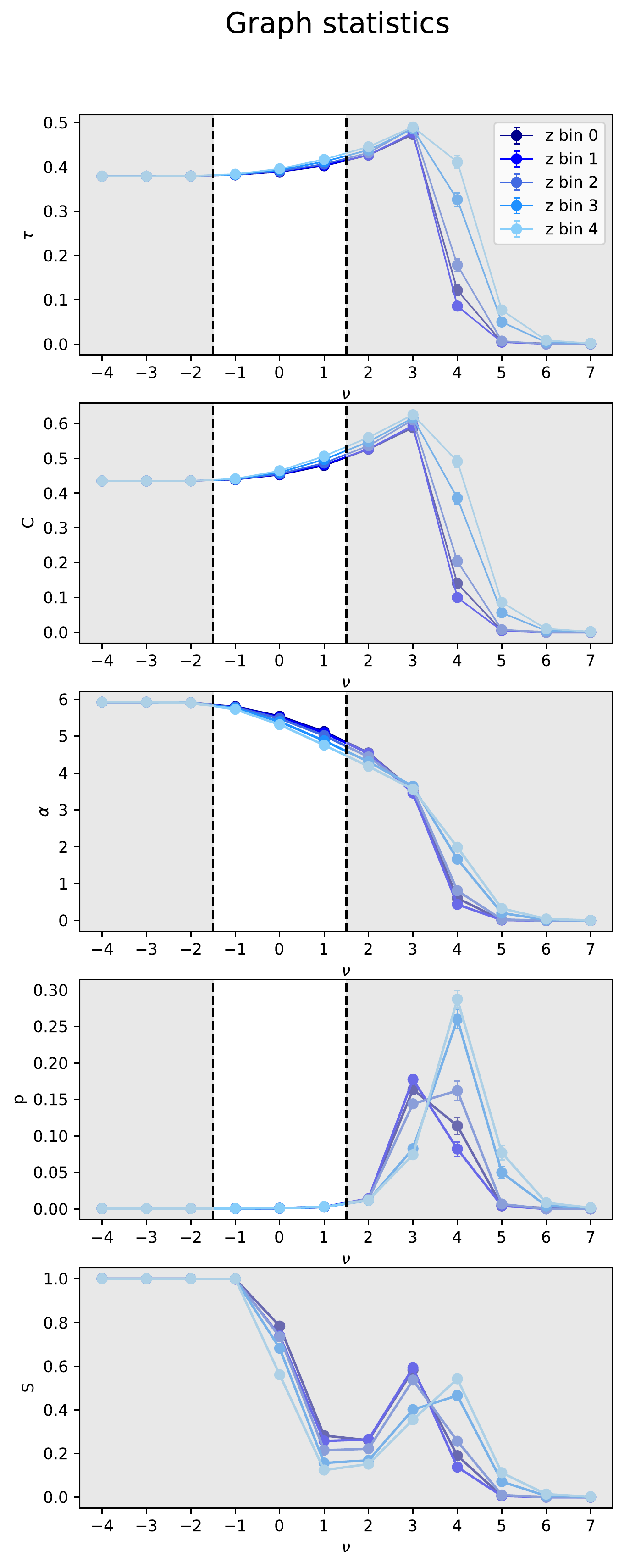}
 \caption{Graph statistics calculated on the maps obtained from the simulation corresponding to \citetalias{planck2018} parameters, averaged over $500$ maps, for all redshift bins, with smoothing scale $\theta_{s}=2\arcminute$. The error bars correspond to the standard deviation divided by the square root of the number of maps. The dashed black lines contain the values selected for the training, the grayed out regions indicate the discarded values.}
\label{fig:graph}
\end{figure}

In Fig.\,\ref{fig:graph}, we show the graph statistics estimated from the simulated maps with \citetalias{planck2018} parameters, averaged over $500$ maps, as a function of the threshold, with $\nu \in [-4,7]$ and $\Delta \nu = 1$, for each redshift bin, and with smoothing scale $\theta_{s}=2\arcminute$. Again, as in Fig.\,\ref{fig:mink} and Fig.\,\ref{fig:betti}, the dashed black lines delimit the values that were selected for the training phase, and the grayed out regions correspond to the discarded values, following the variance ratio criterion. As we can see, we used the values of $\tau$, $C$, $\alpha$, and $p$ in the range $\nu \in [-1,1]$, while $S$ was entirely discarded. We notice that $\tau$ and $C$ have a very similar behavior, as expected considering that $C$ is the vertex-wise version of $\tau$, and that they both increase with $\nu$ and with $z$, meaning that the clustering of the structures in the maps increases both globally and locally as a function of the redshift and of the threshold. On the other hand, $\alpha$, $p$, and $S$ tend to decrease with the redshift, so that while the structures tend to cluster more, they also get smaller. Regarding the behavior with the threshold, while $S$ is too noisy and $p$ is more or less constant in the range considered, $\alpha$ decreases. Therefore, as expected, structures get smaller for high signal-to-noise ratios. Again, for smoothing scales $\theta_{s}=\{4\arcminute, 6\arcminute\}$, the qualitative behavior of the different graph statistics remains unchanged but their absolute value slightly decreases. In fact, for the simulation corresponding to \citetalias{planck2018} parameters, we measured a difference of less than $\sim 15\%$, between the values obtained with $\theta_{s}=2\arcminute$ and with $\theta_{s}=6\arcminute$.

\subsection{Training and test samples}\label{subsec:datasets}

In order to create the final dataset, we collect the measurements of all the higher order estimators described so far. Including the $k_3$ and $k_4$ HOM, the three MFs in the threshold range $\nu \in [-1,1]$, the two Betti numbers, in the range $\nu \in [0,2]$ for $\beta_{0}$ and in the range $\nu \in [-2,0]$ for $\beta_{1}$, and the $\tau$, $C$, $\alpha$, and $p$ graph statistics in the range $\nu \in [-1,1]$, we obtained $29$ measurements for each redshift bin, making a total of $145$ measurements for each simulation. 

Because the measurements on a single map are dominated by the noise, we need to increase the signal-to-noise ratio of the estimators by averaging them over multiple maps. In order to investigate whether we could obtain a better performance from a bigger but noisier dataset or from a smaller dataset with a higher signal-to-noise ratio, we decided to average each measurement over $100$, $300$, and all $500$ maps belonging to the same simulation, obtaining three versions of the dataset. In the first version (hereafter AVG100), we averaged each of the $145$ measurements over $100$ maps, corresponding to a total area of $2500\,\rm{deg}^2$. Because we have $500$ maps for each simulation, we obtained five realizations of the set of $145$ measurements for each combination of the cosmological parameters. Therefore, the dataset passed from \num{750000} to \num{7500} independent realizations. In the second version (hereafter AVG300), we averaged each of the $145$ measurements over $300$ maps, corresponding to a total area of $7500\,\rm{deg}^2$, obtaining just one realization for each cosmological model. This means that there are $200$ maps, among the $500$ maps per simulation, that we did not use. This dataset passed from \num{750000} to \num{1500} independent realizations. Finally, in the third version (hereafter AVG500), we averaged each of the $145$ measurements over $500$ maps, corresponding to a total area of $12500\,\rm{deg}^2$, therefore using all the maps available for each simulation, and obtaining again just one realization of the set of measurements for each combination of the cosmological parameters. This dataset too passed from from \num{750000} to \num{1500} independent realizations. A summary of the three datasets is given in Table\;\ref{tab:dataset_summary}.

\begin{table*}
\begin{center}
\begin{tabular}{lllll}
\hline \hline
dataset & no. of maps per average & no. of realizations per cosmology  & total no. of realizations & equivalent area\\
\hline
AVG100 & 100 &  5 & 7500 & 2500\\
AVG300 & 300 &  1 & 1500 & 7500\\
AVG500 & 500 &  1 & 1500 & 12500\\
\hline\hline
\end{tabular}
\end{center}
\caption{Summary of the three datasets. Starting from \num{750000} independent realizations ($500$ maps for each of the $1500$ cosmologies): in AVG100 each feature is averaged over $100$, corresponding to a total area of $2500\,\rm{deg}^2$, making five realizations for each cosmology and a total of \num{7500} realizations; in AVG300 each feature is averaged over $300$, corresponding to a total area of $7500\,\rm{deg}^2$, making one realization for each cosmology and a total of \num{1500} realizations; in AVG500 each feature is averaged over $500$, corresponding to a total area of $12500\,\rm{deg}^2$, making one realization for each cosmology and a total of \num{1500} realizations.}
\label{tab:dataset_summary}
\end{table*}

Each of the three datasets is then divided into a training set, consisting of 80\% of the respective original dataset, and a test set obtained with the remaining 20\%. Hereafter we will refer to the estimator measurements in the datasets as \textit{features}, and to the corresponding cosmological parameters as \textit{labels}. We repeated this procedure for each smoothing scale, obtaining the three datasets, AVG100, AVG300, and AVG500 for $\theta_{s}=2\arcminute$, and the datasets AVG100 and AVG300 for $\theta_{s}=\{4\arcminute, 6\arcminute\}$. Considering that the application of Gaussian smoothing degrades part of the information contained in the shear maps, we expect to obtain progressively worse results for increasing smoothing scale. For this reason and to save computational time, we chose not to apply the smoothing scales $\theta_{s}=\{4\arcminute, 6\arcminute\}$ to the AVG500 dataset, deeming exhaustive the comparison of the results coming from the AVG100 and AVG300 dataset with those smoothing.  

\begin{figure*}
\centering
\includegraphics[scale=.60]{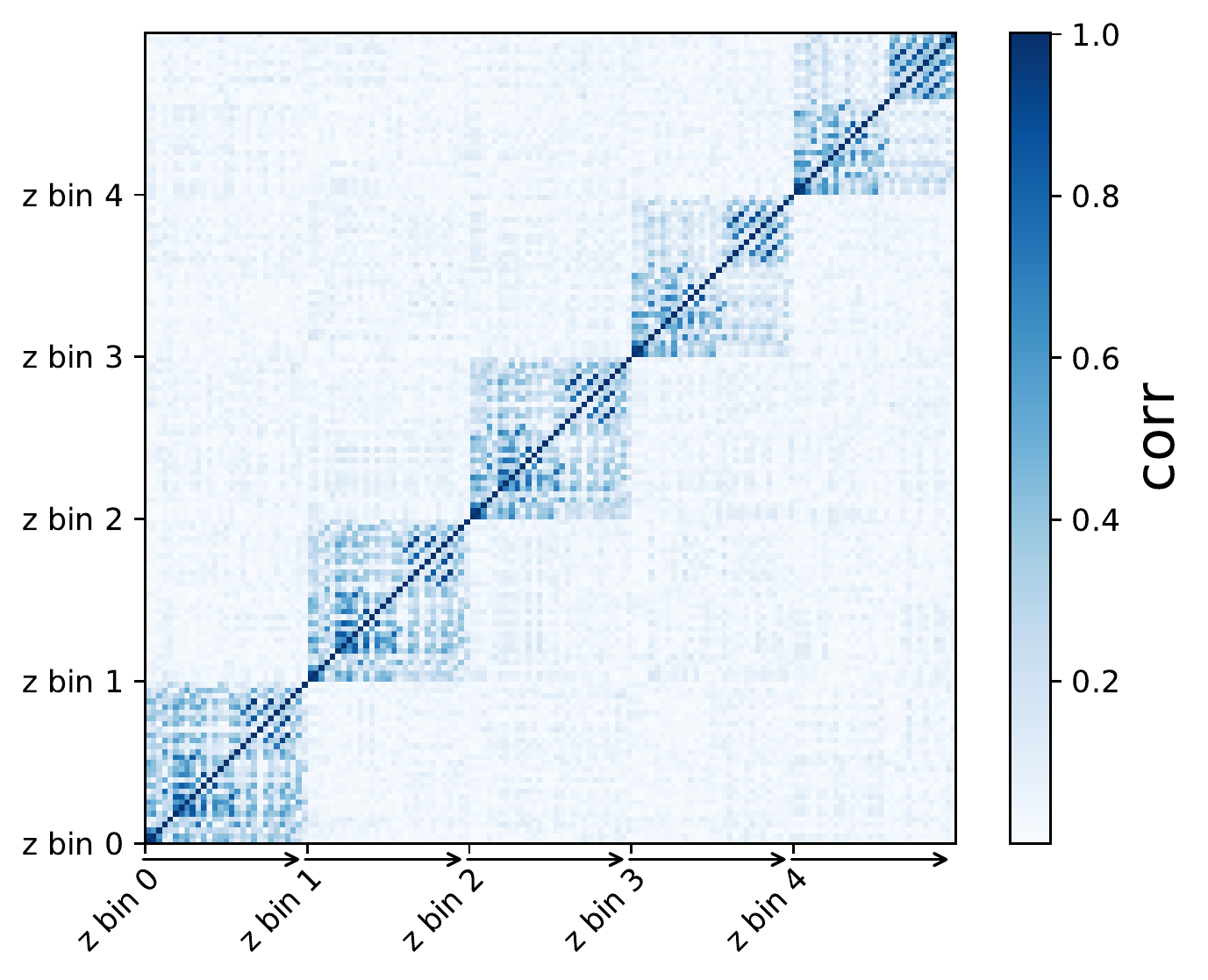}
\includegraphics[scale=.60]{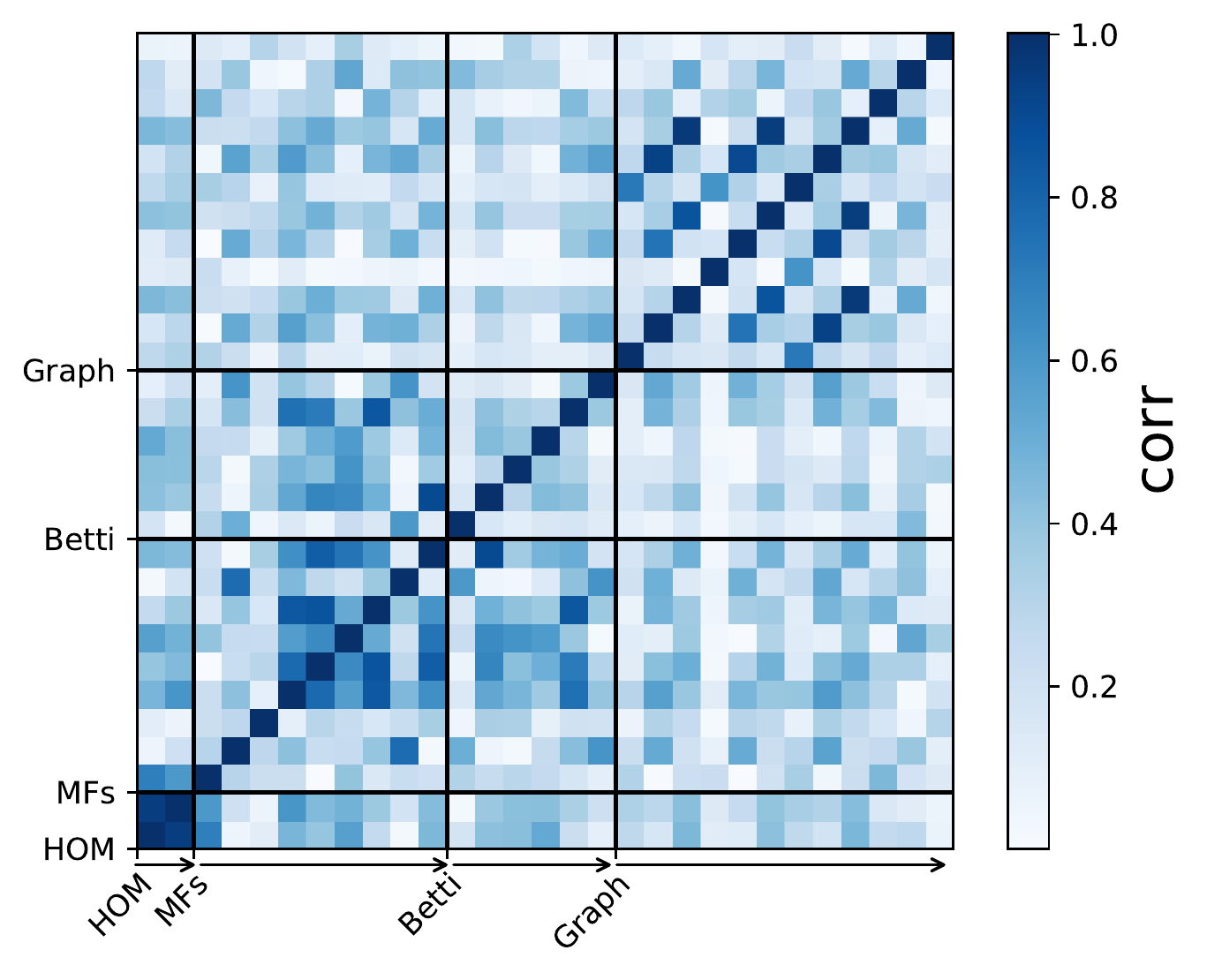}
\includegraphics[scale=.60]{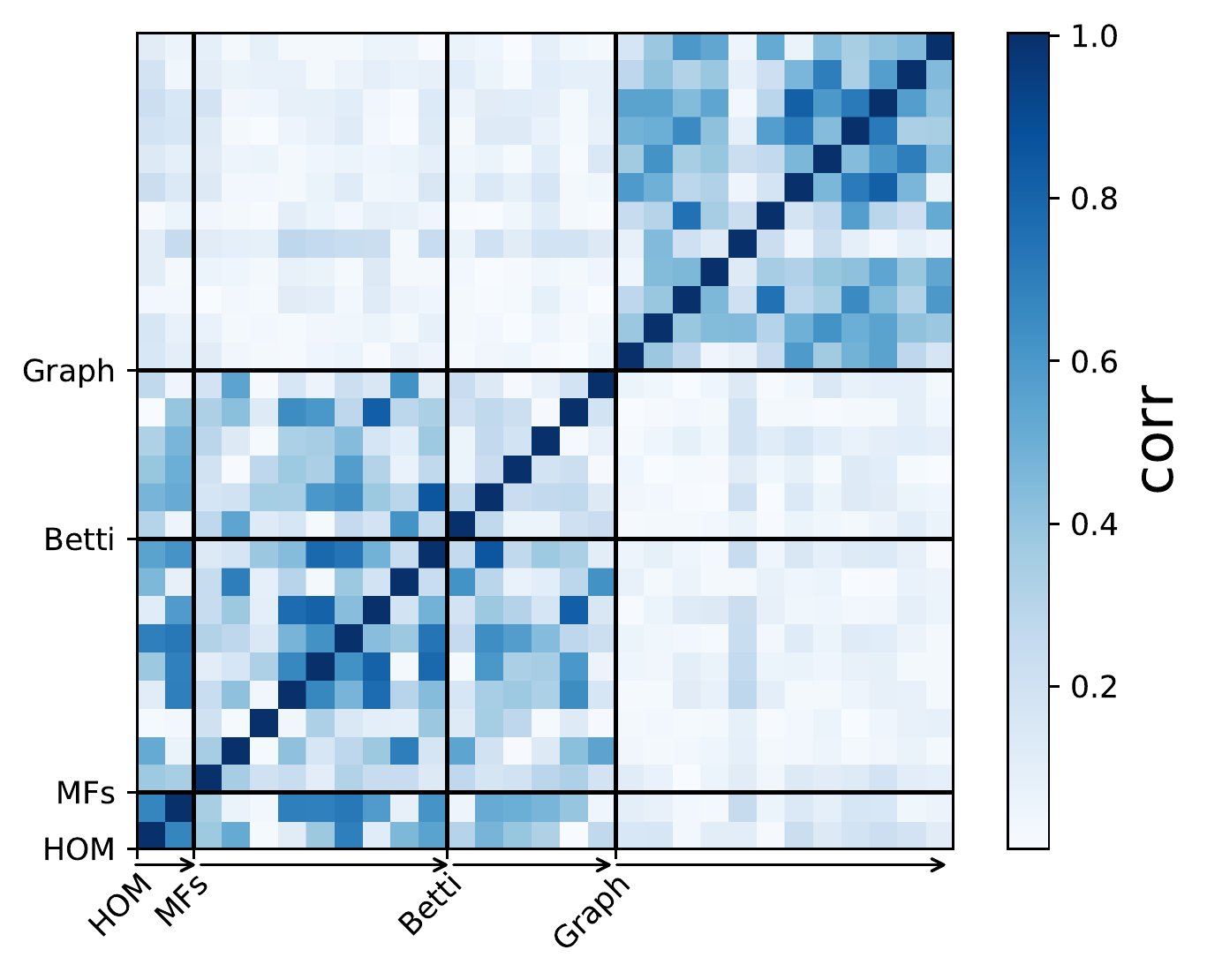}
\includegraphics[scale=.60]{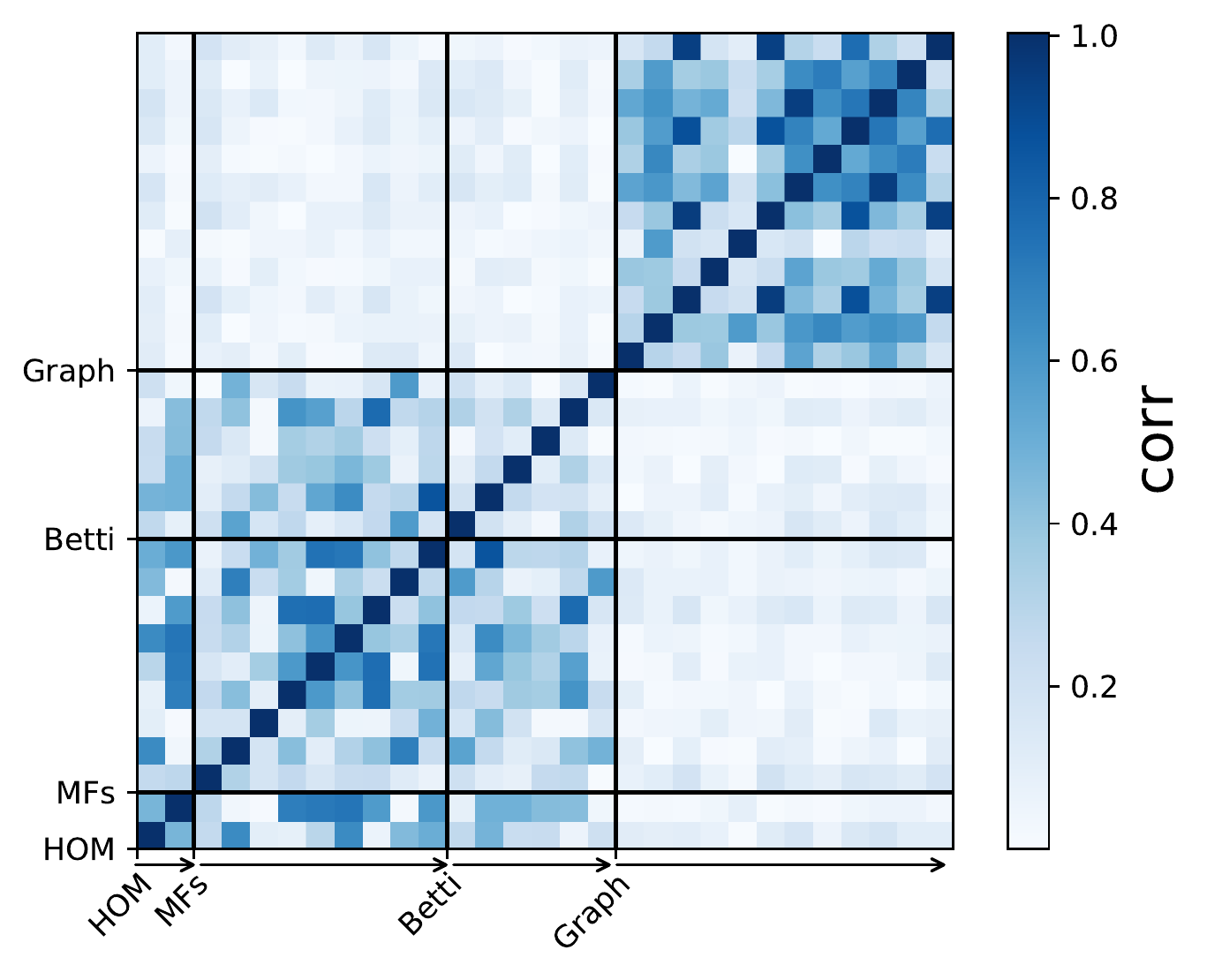}
\caption{\textit{Top Left}: Covariance matrix obtained using all $29$ features for each redshift bin, measured on the $500$ maps from the simulation with \citetalias{planck2018} parameters, with smoothing $\theta_{s}=2\arcminute$. \textit{Top Right}: Same as in the top left panel but for the first redshift bin only. \textit{Bottom Left}: Same as in the top right panel but measured on $300$ maps for smoothing $\theta_{s}=4\arcminute$. \textit{Bottom Right}: Same as in the top right panel but measured on $300$ maps for smoothing $\theta_{s}=6\arcminute$. We notice that within the same simulation, the correlation between features at different redshift bins is quite small due to the adopted binning, while for a given redshift bin some features appear to be more correlated than others. The correlation between the graph statistics and the rest of the estimators further decreases with increasing smoothing scale.}
\label{fig:cov}
\end{figure*}

A comment is in order here about the choice of the map size. Having a side length of $5\,\rm{deg}$ only makes us confident that the flat sky approximation can be used, which is useful given that a full sky treatment of some of the above statistics is not available. Such small maps are, however, likely to be affected by cosmic variance which is a further motivation to average over a large number of them. In a realistic application, this can be done splitting the full survey area in $5 \times 5 \, \rm{deg}^2$ non overlapping maps. This would demand an area of $(2500, 7500, 12500)\,\rm{deg}^2$ in order to create the AVG100, AVG300, AVG500 datasets. Among current ongoing Stage III surveys, DES is compliant with our requirements for AVG100 since Y3 and Y5 data releases will cover $5000\,\rm{deg}^2$. On the contrary, Stage IV surveys will be needed for AVG300 and AVG500 given the large area required. In particular, both Euclid ($15000\,\rm{deg}^2$) and LSST ($18000\,\rm{deg}^2$) will cover enough area for both cases. The methods we are presenting is therefore designed to fully exploit the potentiality of Stage IV surveys.

In Fig.\ref{fig:cov}, we show the correlation between the different features using the $500$ maps from the simulation with \citetalias{planck2018} parameters ($300$ maps for $\theta_{s}=\{4\arcminute,6\arcminute$\}). In the upper left panel, we find the total covariance matrix for smoothing $\theta_{s}=2\arcminute$, which includes all the selected features and all redshift bins, and, in the upper right panel, a zoom on the first redshift bin. We can observe that, within the same simulation, the correlation between features at different redshift bins is quite small due to the adopted binning. On the other hand, for a given redshift bin some features appear to be more correlated than others. In particular, we notice higher correlations between the HOM and the MFs and between the MFs and the Betti numbers, while the graph statistics appear to have slightly lower correlations with the other set of estimators. Such a result is not fully unexpected. At a lowest order, the MFs can be expressed as a perturbative series whose coefficients are related to the generalized moments whose analytical expression is quite similar to that for the HOM. This is telling us that MFs are indeed related to the moments of the distribution so that a correlation can be anticipated. Similarly, the Betti numbers are known to be a generalization of the MFs which explains why they turn out to be correlated with them. On the contrary, the relation between graph statistics and the other estimators has not been investigated up to now so that the lack of correlation we find is an interesting novel property. In the bottom panels, we show again the zoomed covariance matrix of the first redshift bin but for different smoothing scales, $\theta_{s}=4\arcminute$ on the left, and $\theta_{s}=6\arcminute$ on the right. We notice that the correlation between the Betti numbers at different thresholds decreases, while the correlation of the graph statistics increases. On the other hand, the correlation between the graph statistics and the rest of the estimators further decreases, showing a decoupling into two sets of estimators.

\section{Model selection}\label{sec:model_sel}

Using the training and the test sets, obtained as explained in the previous section, we compared different machine learning algorithms in order to establish the model that best describes the relation between the features that we measured on the shear maps and the cosmological parameters.

Explaining the inner workings of the different algorithms and the particular implementations that we used is beyond the scope of this paper. We nevertheless briefly outline the methods used in Appendix \ref{app:models}, referring the interested reader to the specific resources therein for further details. The algorithms that we tested are linear regression, Ridge regression, Kernel Ridge regression, Bayesian Ridge regression, Lasso regression, Support Vector Machine, K Nearest Neighbors, Gaussian Processes, Decision Tree, Random Forests, and Gradient Boosting.    

\begin{table*}
\begin{center}
\begin{tabular}{llllllllllll}
\hline \hline
 & linear & ridge & k. ridge & b. ridge & lasso & SVM & KNN & GP & DT & RF & GB \\
\hline
$\rm{H_0}$ & 0.16 & 0.17 & 0.17 & 0.17 & 0.18 & 0.16 & - & 0.16 & 0.05 & 0.15 & 0.16 \\
$\omega_{\rm{b}}$ & - & - & - & - & - & - & - & - & - & - & - \\
$\Omega_{\rm{M}}$ & 0.61 & 0.61 & 0.61 & 0.61 & 0.61 & - & 0.31 & 0.61 & 0.19 & 0.49 & 0.55 \\
$\Omega_\Lambda$ & 0.61 & 0.61 & 0.61 & 0.61 & 0.61 & - & 0.31 & 0.60 & 0.19 & 0.49 & 0.55 \\
$\rm{w_0}$ & 0.65 & 0.65 & 0.65 & 0.65 & 0.65 & 0.57 & 0.49 & 0.65 & 0.42 & 0.59 & 0.60 \\
$\rm{n_s}$ & 0.16 & 0.16 & 0.16 & 0.16 & 0.16 & - & - & 0.16 & 0.02 & 0.10 & 0.11 \\
$\sigma_8$ & 0.56 & 0.56 & 0.56 & 0.56 & 0.56 & - & 0.30 & 0.56 & 0.25 & 0.48 & 0.49 \\
\hline\hline
\end{tabular}
\end{center}
\caption{$\rm{R}^2$ scores of the best model on the test set, for linear regression, Ridge regression, Kernel Ridge regression, Bayesian Ridge regression, Lasso regression, Support Vector Machine (SVM), K Nearest Neighbors (KNN), Gaussian Processes (GP), Decision Tree (DT), Random Forests (RF), and Gradient Boosting (GB), for each cosmological parameter. The dash represents a negative value of the score. The dataset used is the AVG100 with $\theta_{s}=2\arcminute$. All models tending to the linear regression obtained the same score, with the exception of Support Vector Machine. Overall the best score is obtained with the linear regression.}
\label{tab:model_sel}
\end{table*}

We used the python Scikit-learn \citep{sklearn} library implementation of all the listed algorithms. A limitation of the Scikit-learn implementation of the majority of the methods used is the impossibility of training a model to predict the entire set of labels at once. In fact, even if the labels were chosen randomly and independently, making them uncorrelated (with the exception of $\Omega_{\rm{M}}$ and $\Omega_\Lambda$, which are linked by the assumption of a flat universe, $\Omega_\Lambda = 1-\Omega_{\rm{M}}$), they act simultaneously on the maps and some features could be sensitive to particular parameter combinations. This problem could have been solved varying one cosmological parameter at a time for each simulation. The downside is that the computational time needed to generate the same amount of simulations would be multiplied by the number of cosmological parameters that we are interested in and, more importantly, this approach would produce a training set that would not be representative of observations. In fact, making vary one parameter at a time requires to fix the remaining parameters to some value that, with observations, we do not know {\it a priori}. Therefore we decided to perform the training separately for each cosmological parameter, and considered the effect of the variation of the remaining parameters as additional noise on the features. This means that we trained seven different machine learning models for each algorithm and used them to predict the respective cosmological parameters.

We performed $3$-fold cross-validation to choose the values of the hyperparameters that determine the best model for each method. We evaluated the performance of each model using the $\rm{R}^2$ score, defined as 

\begin{equation}
\begin{aligned}
\rm{R}^2 = & (1-\rm{RSS}/\rm{TSS}) \\
\rm{RSS} = & \sum{\left( y_{\rm{true}}-y_{\rm{pred}} \right)^2} \\
\rm{TSS} = & \sum{\left( y_{\rm{true}}-\left< y_{\rm{true}} \right> \right)^2}
\end{aligned}
\label{eq:r2}
\end{equation}

where $\rm{RSS}$ is the residual sum of squares, $\rm{TSS}$ is the total sum of squares, $y_{\rm{true}}$ are the true labels, and $y_{\rm{pred}}$ are the predicted labels. With this definition, the best score is 1 and a constant model that always predicts the expected value of y, disregarding the input features, would give a score of 0. The score can also be negative, because a model can be arbitrarily worse than the constant model.

All the penalized models (i.e. Ridge, Kernel Ridge, Bayesian Ridge, Lasso, and Support Vector Machine) obtained the best score with a small value of the penalty hyperparameter ($\alpha \leq 10^{-4}$) tending therefore to a simple linear regression model. All the models employing a kernel (i.e. Kernel Ridge, Support Vector Machine, and Gaussian Processes) gave the best performance when using a linear kernel, compared to the other kernels that we tested (i.e. polynomial of degree 2 and 3, RBF, sigmoid, Matern, and rational quadratic kernels from the Scikit-learn library), again making the models tend to a linear regression.

In Table\,\ref{tab:model_sel}, we show the score obtained on the test set for the best model for each method and for each cosmological parameter, with the dash indicating a negative score. The results were obtained using the AVG100 dataset with $\theta_{s}=2\arcminute$. As expected, we can see that all the models tending to the linear model obtained the same score, with the exception of Support Vector Machine that uses a different minimization function compared to linear, Ridge, Kernel Ridge, Bayesian Ridge, Lasso, and Gaussian Processes. Overall the best score is obtained with the linear regression, followed by Gaussian Processes and Random Forests, while K Nearest Neighbors, Decision Tree, and Support Vector Machine perform progressively poorly. The fact that linear regression performs better than the other models can be explained considering the small interval of variation of each cosmological parameter. We are looking at a very zoomed in region of the hyperspace spanned by the features so that locally the relation with the cosmological parameters tends to linearity. The parameter that is best predicted is $\rm{w_0}$ with a promising score of $0.65$, with $\Omega_{\rm{M}}$ and $\Omega_\Lambda$ coming next with a still fairly good score of $0.61$. We obtained a slightly lower score of $0.56$ for $\sigma_8$, while $\rm{H_0}$, and $\rm{n_s}$ obtained a much lower scores, and $\omega_{\rm{b}}$ cannot be predicted at all, having only negative scores for all models. The scores that we obtained confirm the ability of WL to probe certain parameters more than others as was expected from previous cosmic shear studies \citep[e.g. ][]{takada2004,munshi2008} in which $\Omega_{\rm{M}}$ and $\sigma_8$ were tightly constrained and, using external CMB measurements of $\rm{H_0}$, $\rm{n_s}$, and $\omega_{\rm{b}}$, it was possible to also improve the constraints on the dark energy equation of state parameters. Increasing the smoothing scale to $\theta_{s}=\{4\arcminute,6\arcminute\}$, we obtain the same qualitative results in terms of the performance of one model with respect to another but overall progressively lower scores. We investigate in more detail the effect of the smoothing in the next Section.

\section{Prediction of the cosmological parameters}\label{sec:prediction}

Considering the results outlined in the previous section, we decided to retain for the rest of this work only the best performing model, which resulted to be linear regression. Using this model we want to study the impact that the size of the training set (i.e. the number of independent realizations), the signal-to-noise ratio of the features (i.e. the number of maps used to average the features), and the smoothing have on the results. We also want to determine which is the best predicted parameter and verify if the results are consistent using the different training sets.

\begin{table}[!h]
\begin{center}
\begin{tabular}{llll}
\hline \hline
 & AVG100, $\theta_{s}=2\arcminute$ & AVG300, $\theta_{s}=2\arcminute$ & AVG500, $\theta_{s}=2\arcminute$ \\
\hline
$\rm{H_0}$        & 0.16 & 0.09 & 0.03 \\
$\omega_{\rm{b}}$ & -    & -  & -      \\
$\Omega_{\rm{M}}$ & 0.61 & 0.64 & 0.70 \\
$\Omega_\Lambda$  & 0.61 & 0.64 & 0.70 \\
$\rm{w_0}$        & 0.65 & 0.75 & 0.81 \\
$\rm{n_s}$        & 0.16 & 0.23 & 0.32 \\
$\sigma_8$        & 0.56 & 0.73 & 0.80 \\  
\hline \hline 
\end{tabular}
\end{center}
\caption{$\rm{R}^2$ score for the three datasets, AVG100, AVG300, and AVG500 with smoothing scale $\theta_{s}=2\arcminute$, for each cosmological parameter. The dash represents a negative value of the score. As expected, increasing the number of maps per simulation improves the performance of the algorithm.}
\label{tab:lin_reg_smooth2}

\begin{center}
\begin{tabular}{lll}
\hline \hline
 & AVG100, $\theta_{s}=4\arcminute$ & AVG300, $\theta_{s}=4\arcminute$ \\
\hline
$\rm{H_0}$        & 0.16 & -     \\
$\omega_{\rm{b}}$ & -    & -     \\
$\Omega_{\rm{M}}$ & 0.55 & 0.60  \\
$\Omega_\Lambda$  & 0.55 & 0.60  \\
$\rm{w_0}$        & 0.62 & 0.68  \\
$\rm{n_s}$        & 0.06 & -     \\
$\sigma_8$        & 0.52 & 0.64  \\  
\hline \hline 
\end{tabular}
\end{center}
\caption{Same as in Table\ref{tab:lin_reg_smooth2} but for the AVG100 and AVG300 datasets with smoothing scale $\theta_{s}=4\arcminute$.}
\label{tab:lin_reg_smooth4}

\begin{center}
\begin{tabular}{lll}
\hline \hline
 & AVG100, $\theta_{s}=6\arcminute$ & AVG300, $\theta_{s}=6\arcminute$ \\
\hline
$\rm{H_0}$        & 0.18 & 0.06 \\
$\omega_{\rm{b}}$ & -    & -    \\
$\Omega_{\rm{M}}$ & 0.55 & 0.61 \\
$\Omega_\Lambda$  & 0.55 & 0.61 \\
$\rm{w_0}$        & 0.60 & 0.60 \\
$\rm{n_s}$        & 0.04 & -    \\
$\sigma_8$        & 0.47 & 0.52 \\  
\hline \hline 
\end{tabular}
\end{center}
\caption{Same as in Table\ref{tab:lin_reg_smooth2} but for the AVG100 and AVG300 datasets with smoothing scale $\theta_{s}=6\arcminute$. Comparing this results with those in Table\,\ref{tab:lin_reg_smooth2} and Table\,\ref{tab:lin_reg_smooth4}, we observe that overall the score decreases with increasing smoothing scale.}
\label{tab:lin_reg_smooth6}
\end{table}

\begin{figure*}
\centering
\includegraphics[scale=.50]{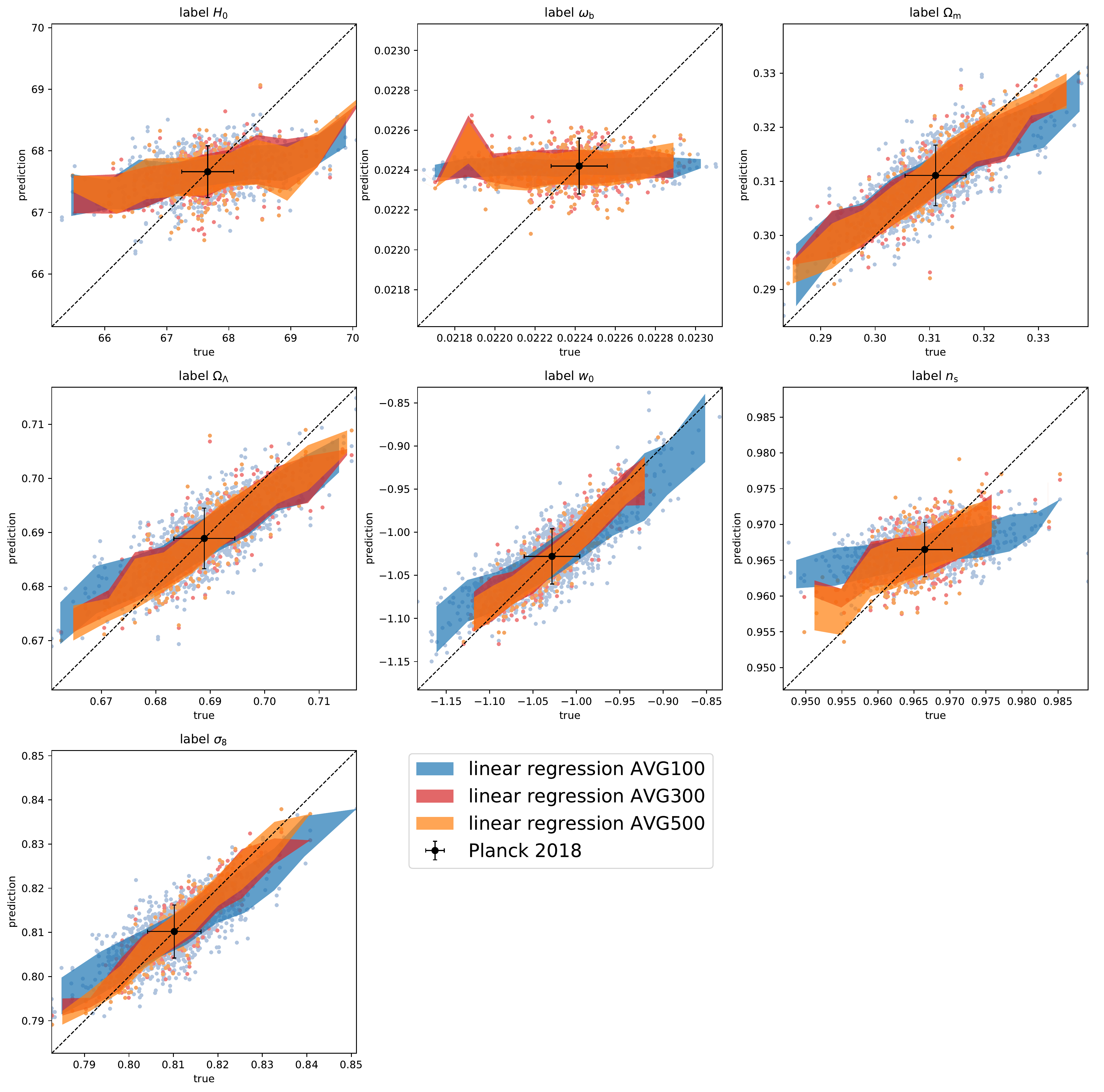}
\caption{Performance on the test set of the linear regression model for each cosmological parameter, and for the three datasets, AVG100, AVG300, and AVG500, with $\theta_{s}=2\arcminute$. The dots represent individual predictions. The shaded regions delimit the $1~\sigma$ interval obtained dividing the test sample into $10$ bins of the true label values and calculating the mean and standard deviation of the predictions. We obtain the best predictions with the AVG500 dataset and for the $\Omega_{\rm{M}}$, $\Omega_\Lambda$, $\rm{w_0}$, and $\sigma_8$ parameters.}
\label{fig:lin_reg}
\end{figure*}

\begin{figure*}
\centering
\includegraphics[scale=.50]{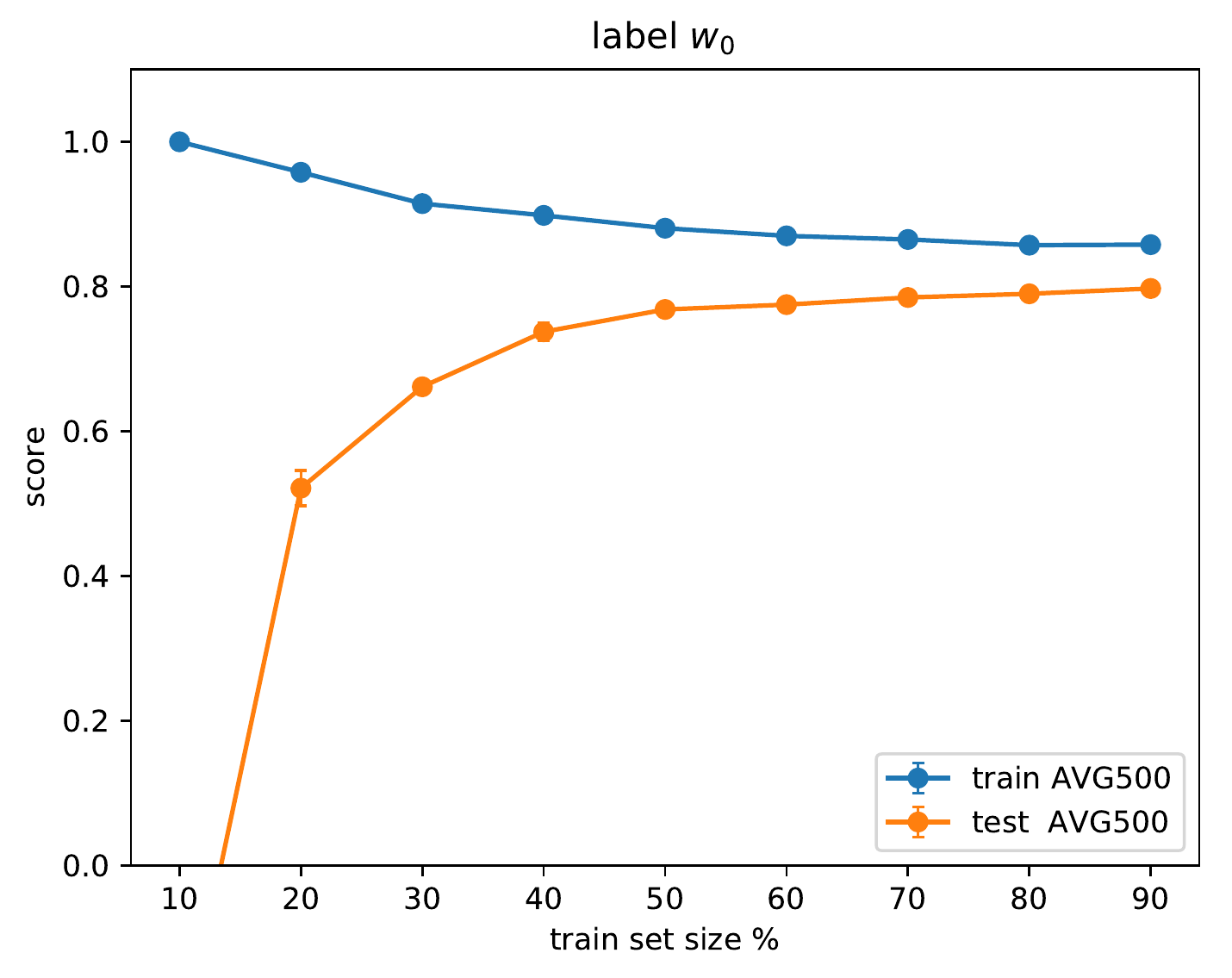}
\includegraphics[scale=.50]{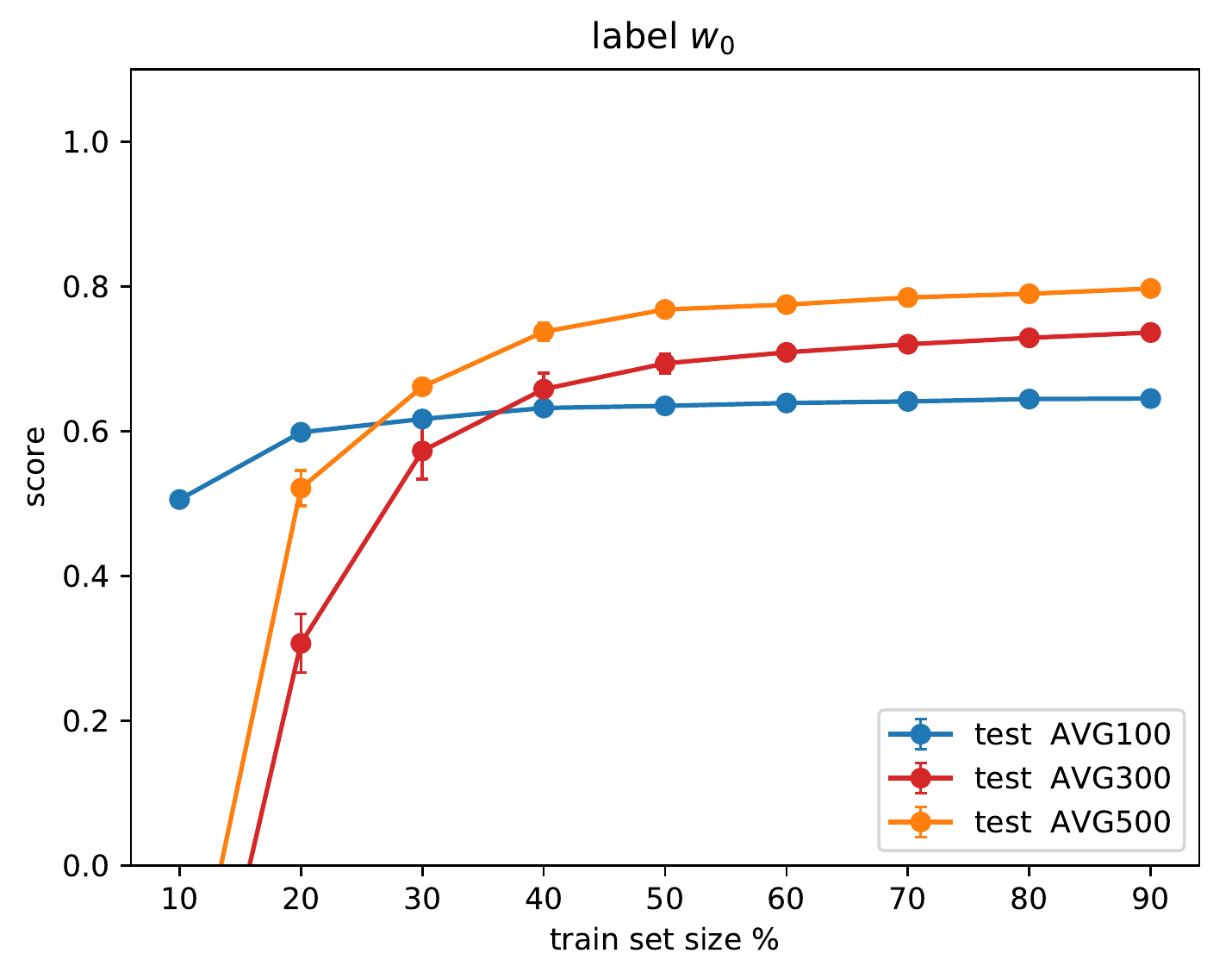}
\includegraphics[scale=.50]{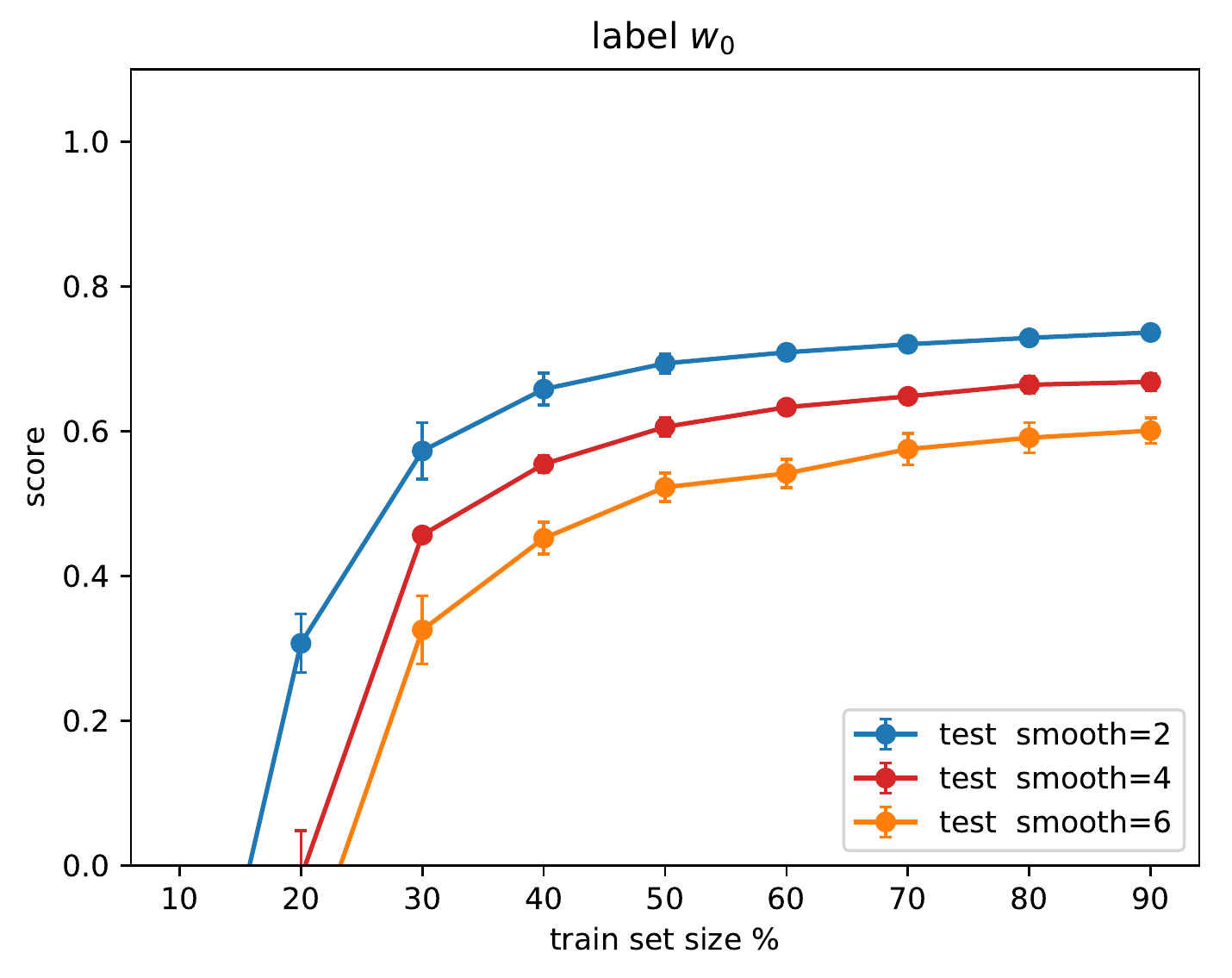}
\includegraphics[scale=.50]{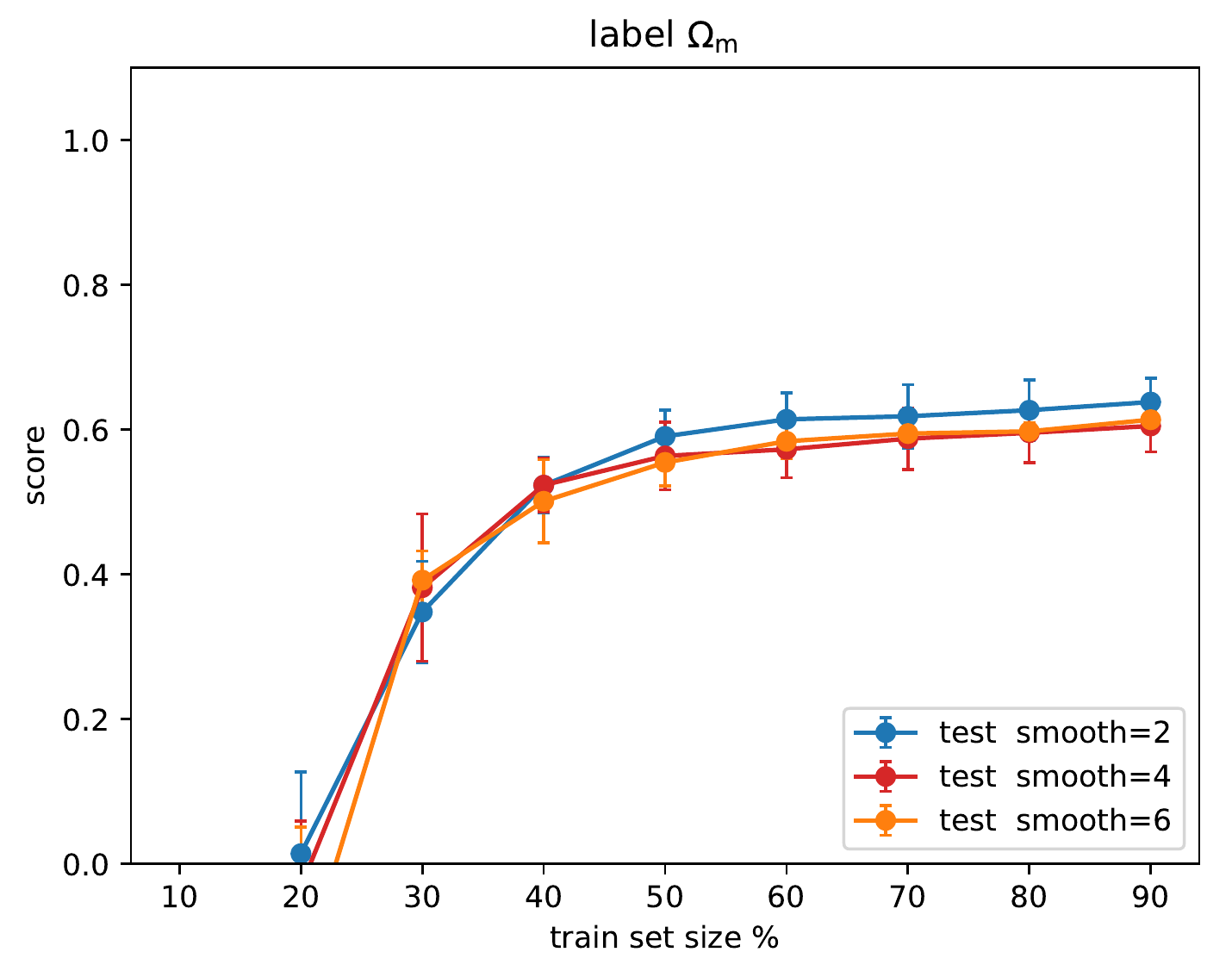}
\caption{\textit{Top left}: Learning curve for the train and the test set of the AVG500 dataset, with smoothing $\theta_{s}=2\arcminute$, for the $\rm{w_0}$ parameter. \textit{Top right}: Learning curve for the test set of the AVG100, AVG300, and AVG500 datasets, with smoothing $\theta_{s}=2\arcminute$, for the $\rm{w_0}$ parameter. \textit{Bottom left}: Learning curve for the test set of the AVG300 dataset, with smoothing $\theta_{s}=\{2\arcminute,4\arcminute,6\arcminute\}$, for the $\rm{w_0}$ parameter. \textit{Bottom right}: Same as in the middle panel but for the $\Omega_{\rm{M}}$ parameter. In all cases we performed 3-fold cross validation.}
\label{fig:learning_curve}
\end{figure*}

We performed the training and the prediction for each cosmological parameter on all three datasets, AVG100, AVG300, and AVG500 with smoothing $\theta_{s}=2\arcminute$ and on the AVG100 and the AVG300 datasets with the additional smoothing scales $\theta_{s}=\{4\arcminute,6\arcminute\}$, in order to compare the results.

In Table\,\ref{tab:lin_reg_smooth2},\ref{tab:lin_reg_smooth4}, and \ref{tab:lin_reg_smooth6}, we show the $\rm{R}^2$ score that we obtained in each case. The dash represents a negative value of the score. In the following, we will only discuss the $\rm{R}^2$ score because we found that, given our dataset, it is the most meaningful statistics to evaluate the model performance. In Appendix \ref{app:mse_mape}, we will also consider two additional statistics, the mean squared error and the mean absolute percentage error.

Starting from Table\,\ref{tab:lin_reg_smooth2}, we compare the results obtained with the AVG100, AVG300, and AVG500 datasets with smoothing $\theta_{s}=2\arcminute$. We can see that while $\rm{H_0}$ and $\omega_{\rm{b}}$ cannot be predicted with any dataset version, the results for the other parameters improve progressively. The best measured parameter is still $\rm{w_0}$ for which we obtained the improved score of $0.75$ using the AVG300 dataset, and the even better score of $0.81$ with the AVG500 dataset, corresponding to a score improvement of $\sim25\%$. The two parameters which obtained the next best score with the AVG100 dataset, $\Omega_{\rm{M}}$ and $\Omega_\Lambda$, both got a smaller but still consistent improvement of $\sim15\%$ in score reaching a value of $0.64$ and $0.70$, with the AVG300 and the AVG500 datasets respectively. The parameters that obtained the largest score improvement are $\sigma_8$ which went from a value of $0.56$ for the AVG100 dataset, to $0.73$ for the AVG300 dataset, reaching the very good score of $0.80$ for the AVG500 dataset with a total score improvement of $\sim43\%$, making it the second best predicted parameter, and $\rm{n_s}$ which improved its score of $\sim100\%$, going from $0.16$ to $0.23$ for the AVG300 dataset, and $0.32$ for the AVG500 dataset.

Passing to Table\,\ref{tab:lin_reg_smooth4}, we can see the results obtained from the AVG100 and the AVG300 datasets with smoothing $\theta_{s}=4\arcminute$ and observe an overall decrease in score. Comparing the AVG100 results with $\theta_{s}=2\arcminute$ and $\theta_{s}=4\arcminute$, we notice that the most affected parameters are $\Omega_{\rm{M}}$ and $\Omega_\Lambda$, which got their score degraded of $\sim10\%$. On the other hand, the same comparison for the AVG300 datasets shows that we obtained the greatest decrease in score for $\rm{w_0}$ and $\sigma_8$, corresponding to $\sim10\%$ and $\sim14\%$, respectively. We also remark that $\rm{n_s}$ cannot be predicted with $\theta_{s}=4\arcminute$ with neither the AVG100 nor the AVG300 dataset.

In Table\,\ref{tab:lin_reg_smooth6}, we show the results obtained from the AVG100 and the AVG300 datasets with smoothing $\theta_{s}=6\arcminute$. We notice that while the scores of $\Omega_{\rm{M}}$ and $\Omega_\Lambda$ are stable compared to the results presented in Table\,\ref{tab:lin_reg_smooth4}, the score of $\rm{w_0}$ and $\sigma_8$ further decreased of $\sim10-20\%$. Again, $\rm{n_s}$ cannot be measured with this smoothing scale.

Considering the behavior of the score for the different cosmological parameters that we observed with increasing smoothing, we can conclude that $\Omega_{\rm{M}}$ and $\Omega_\Lambda$ are less affected by the degradation of the spatial resolution and of the information that we can measure on the shear maps compared to $\rm{w_0}$, $\sigma_8$, and $\rm{n_s}$. In fact, as we will see in Section\,\ref{sec:feat_imp}, the features that most contribute to the measurement of $\Omega_{\rm{M}}$ and $\Omega_\Lambda$ are the graph statistics, which resulted to be much less sensitive to the smoothing scale compared to the other estimators. We also remark that the decoupling between the graph statistics estimators and the HOM, MFs, and Betti numbers that we observed in the correlation matrix presented in Fig.\ref{fig:cov} for increasing smoothing, did not help the algorithm to extract additional information, being probably counterbalanced by the increased correlation inside the set of graph statistics estimators.

Fig.\,\ref{fig:lin_reg} shows the true labels versus the predicted labels, for each cosmological parameter, using the AVG100, the AVG300, and the AVG500 datasets, with smoothing $\theta_{s}=2\arcminute$. The dots represent individual predictions while the shaded areas correspond to the $1~\sigma$ region, obtained dividing the test sample into $10$ bins of the true label values and calculating the mean and standard deviation of the predictions inside each bin. The more the colored region for each given parameter aligns along the dashed black diagonal, the better will be the prediction obtained with such model. The black dots are the values of \citetalias{planck2018} parameters with the respective error, reported also on the $y-$axis as reference. We can see that the alignment along the diagonal improves passing from the blue, to the red, and to the orange regions, corresponding to the AVG100, the AVG300, and the AVG500 datasets, and that we obtain the best results for the $\Omega_{\rm{M}}$, $\Omega_\Lambda$, $\rm{w_0}$, and $\sigma_8$ parameters, confirming what shown in Table\,\ref{tab:lin_reg_smooth2},\ref{tab:lin_reg_smooth4}, and \ref{tab:lin_reg_smooth6}.

In the top left panel of Fig.\,\ref{fig:learning_curve}, we show the learning curve for the $\rm{w_0}$ parameter, which corresponds to the score as a function of the train set size (as percentage values of the complete dataset size, reported in Table\,\ref{tab:dataset_summary}), for the training and the test set of the AVG500 dataset, with smoothing $\theta_{s}=2\arcminute$. We performed 3-fold cross validation and plotted the mean values, with the error bars corresponding to the standard deviation of the different realizations. We see that as the train set size increases, the train score decreases and the test score increases. For example, using $\sim 20\%$ of the entire dataset for training and the remaining $\sim 80\%$ for testing, results in a training score $>0.9$ and in a test score of $\sim 0.5$. When the training score is much higher than the test score, we are in a situation of overfitting, i.e. the model performs almost perfectly on the data that it has seen before but very poorly on new data. Around a train set size of $50\%$ the two curves start to flatten and come out of the overfitting region. We can verify that for our default choice of a training set size $\gtrsim 80$, our model does not overfit the data because the difference between the training and the test score is small.

In the top right panel of Fig.\,\ref{fig:learning_curve}, we show the learning curve for the $\rm{w_0}$ parameter for the test set of the three datasets, with smoothing $\theta_{s}=2\arcminute$. We notice that the blue curve corresponding to the AVG100 dataset flattens out at a train set size of $\sim 50\%$ and a test score of $\sim 0.6$. The red and the orange curves, representing the AVG300 and the AVG500 datasets, respectively, have a different behavior. We can see that for a train set size $<30\%$ we obtain low or negative scores, while for a train set size $>30\%$ the test set curves show an increasing score that reaches values of $\sim 0.7$ for the AVG300 dataset and $\sim 0.8$ for the AVG500 dataset, yet they do not reach a plateau value. This is due to the smaller size of the AVG300 and the AVG500 datasets, which is one fifth of the AVG100 dataset. We can conclude that the number of maps used to average the features in order to increase the signal-to-noise ratio of the measurements has a big impact on the results and it is proportional to the $\rm{R}^2$ score obtained. On the other hand, while increasing the dataset size for the AVG100 version would not change the results, for the AVG300 and the AVG500 we can envisage a margin of improvement of a few percent. 

In the bottom left and bottom right panels of Fig.\,\ref{fig:learning_curve}, we show again the learning curve for $\rm{w_0}$ and $\Omega_{\rm{M}}$, respectively, but this time we compare the results obtained on the test set using the AVG300 dataset with smoothing $\theta_{s}=\{2\arcminute,4\arcminute,6\arcminute\}$. Looking at the plot for $\rm{w_0}$, we notice the same increasing trend for the three curves as for the AVG300 and AVG500 curves in the top panel for smoothing $\theta_{s}=2\arcminute$. Confirming the results shown in Table\,\ref{tab:lin_reg_smooth4} and \ref{tab:lin_reg_smooth6}, the red and orange curves reach a lower score compared to the blue curve. Even if the margin of improvement of a few percent with increasing dataset size is apparent for all three cases, we remark that the plateau value will be lower for increasing smoothing. From the same analysis for the $\Omega_{\rm{M}}$ plot, we observe that the curves corresponding to smoothing $\theta_{s}=\{4\arcminute,6\arcminute\}$ are more or less superposed and that they reach a score value very close to the blue curve. This consolidates the conclusions that we drew, comparing the results of Table\,\ref{tab:lin_reg_smooth4} and \ref{tab:lin_reg_smooth6} for the different cosmological parameters, on the higher sensitivity of $\rm{w_0}$, $\sigma_8$, and $\rm{n_s}$ to some particular information contained in the shear maps that is degraded by the smoothing, compared to $\Omega_{\rm{M}}$ and $\Omega_{\Lambda}$.

Summarizing, we conclude that overall the model performance increases as a function of the signal-to-noise ratio of the features (or the number of maps used for the averages, i.e. the total area) and decreases as a function of the smoothing scale.

\section{Feature importance}\label{sec:feat_imp}

Once we obtained the predictions of the cosmological parameters, we wanted to investigate their relation with the higher order estimators used to obtain them. In order to estimate which features contribute more to the measurement of each cosmological parameter, we repeated the training using random subsets of the features in the AVG500 dataset, with smoothing $\theta_{s}=2\arcminute$. We then assigned a value to each feature summing the score divided by number of features in the subset, for each random realization, increasing the number of iterations until convergence, for a total of \num{800000} realizations. We used random subsets with sizes between 3 and 15 features. We then normalized the feature contributions so that they assume a value between 0 and 1.

\begin{figure*}
\centering
\includegraphics[scale=.80]{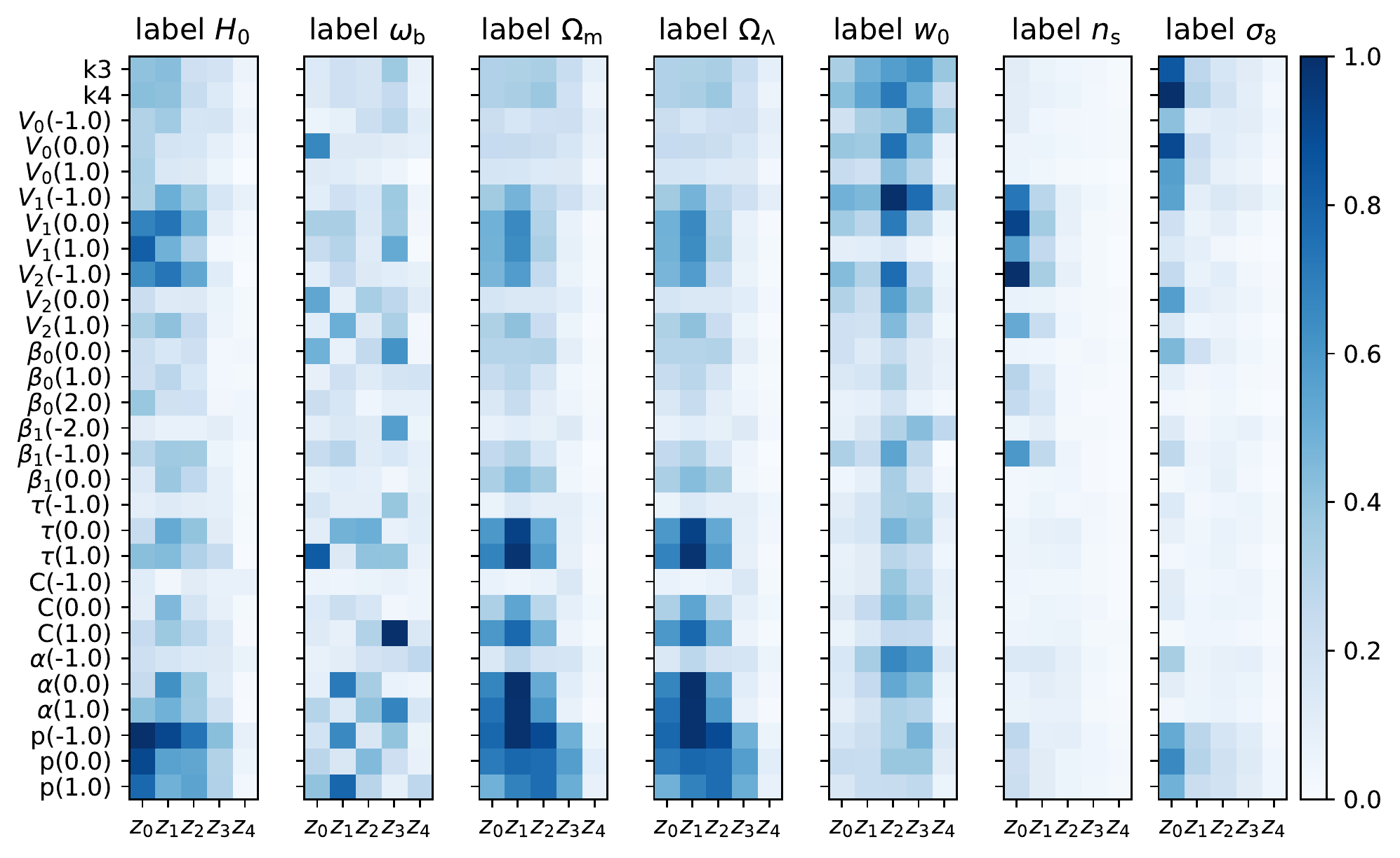}
\caption{Normalized feature importance for each cosmological parameter. Graph statistics have a fundamental importance for the measurement of $\Omega_{\rm{M}}$ and $\Omega_\Lambda$, as do the MFs for $\rm{n_s}$, and the HOM for $\sigma_8$, while $\rm{w_0}$ slightly favors HOM and MFs. Overall the information measured from the different estimators increases with decreasing redshift. From left to right, the $\rm{R^2}$ scores are $\{0.03,-,0.70,0.70,0.81,0.32,0.80\}$, as reported in Table\,\ref{tab:lin_reg_smooth2}.}
\label{fig:feat_imp}
\end{figure*}

In Fig.\,\ref{fig:feat_imp}, we show a color map that indicates the importance of each feature, i.e. the contribution of each feature to the final score, for each cosmological parameter. We remark that, being the feature contributions normalized separately for each color map, they offer a measure of the importance of each feature compared to the others, in the prediction of a given cosmological parameter. When comparing the color maps between them, we have to keep in mind the $\rm{R}^2$ scores for the AVG500 dataset, reported in Table\,\ref{tab:lin_reg_smooth2}, for each cosmological parameter. Color maps for parameters with low  $R^2$ score should be taken with caution, which is the case for $\rm{H_0}$, $\omega_{\rm{b}}$, and $\rm{n_s}$. We nevertheless report them for completeness.

It turns out that the color maps for $\rm{H_0}$, $\Omega_{\rm{M}}$, and $\Omega_{\Lambda}$ show the same qualitative features. For all of them, the HOM, the zeroth-order MF $V_0$, and the Betti numbers give a negligible contribution to the final score which is dominated by the other two MFs and the $(\tau, C, \alpha, p)$ graph statistics. Low\,-\,medium redshift bins are preferred with the bin at $z = 0.9$ for graph statistics playing the major role in determining $\Omega_{\rm{M}}$ and $\Omega_{\Lambda}$.

It is, on the contrary, hard to interpret the color map for $\omega_{\rm{b}}$ with its almost random distribution of colors. This is, however, not surprising given the negative score in every version of the feature dataset used. Again, this could have been anticipated since $\omega_b$ is mainly responsible for modulating the BAO wiggles in the matter power spectrum which are smoothed out by the lensing kernel. As such, WL is not sensible to this parameter no matter which estimators (second or higher order) one relies on.

The color map for $\rm{w_0}$ points at the HOM and MFS as main contributors with no particular preference for a redshift bin, while graph statistics and Betti numbers follow with the medium-high redshift range contributing the most. The need to use all the redshift bins (although with different estimators) is likely related to the need to follow the growth of structures whose evolution is determined by the $\rm{w_0}$ value. Which estimator is best suited depends on the level of non-Gaussianity. At low redshift, the nonlinear collapse of structures enhances the non-Gaussianity of the field which can be quantified by the easy to measure HOM an MFs. On the contrary, at larger $z$, one is approaching the linear regime hence the need for more advanced tools to spot the residual non-Gaussianity.

Moving to $\rm{n_s}$, the color map points at $V_1$, $V_2$, and $\beta_1$ at low redshift as leading contributor with $\beta_0$ and $p$ giving the residual contribution. Remember, however, that $\rm{n_s}$ is poorly determined overall so that the color map is less informative.

Finally, the color map for $\sigma_8$ shows some similarity with the one for ${\rm w_0}$, the main contribution coming from the HOM, followed by the MFs and the $p$ parameter. However, the importance of the parameters now increases as $z$ decreases, with the first redshift bin bringing the majority of the information.

Excluding $\omega_{\rm{b}}$ for its erratic behavior from the rest of this discussion, we can observe that overall the information measured from the different estimators increases with decreasing redshift (dramatically so for $\rm{n_s}$ and $\sigma_8$), with the exception of $\rm{w_0}$. While for $\Omega_{\rm{M}}$, and $\Omega_\Lambda$ the peak is reached at $z=0.9$, the most contributing redshift bin for $\rm{n_s}$ and $\sigma_8$ is clearly $z=0.6$. On the other hand, for $\rm{w_0}$ the importance appears to be more uniform in redshift, with a slight preference for $z=1.2-1.5$. 

While MFs seems to be in some measure sensitive to all the cosmological parameters, overall the Betti numbers appear to contribute the least, with a small exception at some thresholds for $\rm{n_s}$ and $\sigma_8$. Graph statistics have a fundamental importance for the measurement of $\Omega_{\rm{M}}$ and $\Omega_\Lambda$, as do the MFs for $\rm{n_s}$, and the HOM for $\sigma_8$. Once again, $\rm{w_0}$ shows a more uniform behavior also in terms of estimators contribution, slightly favoring HOM and MFs. This explains somehow the decrease in score for the different cosmological parameters with increasing smoothing scale. As we discussed in Section\,\ref{sec:prediction}, the score of $\Omega_{\rm{M}}$ and $\Omega_\Lambda$ is more stable as a function of the smoothing, compared to the score of $\rm{w_0}$, $\sigma_8$, and $\rm{n_s}$, and this behavior is reflected by the most important features for each parameter. In fact, while the HOM, MFs, and Betti numbers change with the smoothing of a factor $\sim 40$, $\sim 30$, and $\sim 10$, respectively, the graph statistics only vary of less than $\sim 15\%$.     

The feature importance can also be interpreted considering the physical meaning of the different estimators. The measurement of $\Omega_{\rm{M}}$, and $\Omega_\Lambda$ parameters is mainly due to $\tau$, $C$, $\alpha$, and p, making these parameters sensitive to overall degree of connectivity or clustering measured on the shear maps. The $\rm{n_s}$ parameter is linked to the information that is contained in the derivatives of the shear field, through $V_1$ and $V_2$, which are in turn connected to the matter power spectrum and bispectrum, while in addition to this, $\sigma_8$ is also related to a greater extent to the variance of the shear field through $V_0$ and to the three- and fourth-point correlation functions, through $k_{3}$ and $k_{4}$. Finally, all of the above contributes to the measurement of $\rm{w_0}$.  

\section{Improving the methodology}\label{sec:discussion}

The main aim of this paper was to present a new methodology to use WL higher order statistics starting from the shear field with no need for the convergence reconstruction and for a theoretical formulation of the relation between the estimators used and the underlying cosmology. The interesting results discussed above are a good reason to reconsidering the limitations of this first step in order to understand how to make the method still more appealing.

A point worth improvement is the realism of the training set of simulations. We have indeed approximated the shear field as lognormal, and used {\tt FLASK} to quickly generate a large set of maps varying the cosmological parameters. Both these aspects can be ameliorated. First, we note that the requirement that the lognormal approximation is a good representation of the shear field has forced us to consider only bins with $z > 0.5$ thus cutting out the low redshift regime, which is the one dominated by the dark energy we want to investigate. Since we have considered only models with constant equation of state, it has not been of paramount importance to investigate where the transition from accelerated to decelerated expansion takes place. Adding $w_a$ to the list of parameters would probably ask for the inclusion of lower redshift bins. Also the fact that \texttt{FLASK} does not make any assumption on the shear higher order moments, making them unreliable for realistic simulations, implies that some of the estimators that we considered could lead to different results on actual observations. Moving beyond {\tt FLASK} is therefore be necessary in order to create a training set which is as similar as possible to the underlying true universe. For this same reason, it is of fundamental importance to adopt the correct source redshift distribution and account for the errors due to photo\,-\,z. Note that both these aspects are survey dependent so that a reliable training exercise can be obtained only with a good knowledge of the survey specifics. Moreover, intrinsic alignment, which in the weak regime linearly adds to the lensing shear, should also be taken into account. Even if intrinsic alignment is a local effect that should not alter the global topology of the maps and it is subdominant at high redshift, it increases the correlation among close redshift bins and therefore might also increase the correlation among features at different $z$, decreasing their constraining power. It also represents an additional source of noise that should be modeled and included in the simulations \citep[e.g.][]{bruderer2016, hildebrandt2017, wei2018, ghosh2020}. Another missing ingredient are baryonic effects, which affects the cosmic shear signal at medium angular scales, where the total matter power spectrum is subjected to a suppression of power up to $\sim 30\%$, and at very small scales, where the power is enhanced because of efficient baryon cooling and star formation in the halo centers. These effects have been modeled both numerically and analytically but different implementations lead to different results. While the general trend is reproduced in most simulations, there is still no agreement on the quantitative level so that more work need to be done to reach a self-consistent treatment of these processes in the cosmological context (e.g. \citealt{harnois2015}; \citealt{chisari2018}; Euclid Collaboration \citeyear{euclid2019}; \citealt{kacprzak2019}; \citealt{schneider2019}; \citealt{schneider2020b,schneider2020a}).

Another aspect worth improving is the range which the cosmological parameters are varied over to generate the simulated maps. Our initial goal was to compare the constraints one can get from WL high order statistics with those from the joint use of CMB and other probes reported in \citetalias{planck2018}. This motivated us to choose the \citetalias{planck2018} values as reference and the corresponding $1 \sigma$ errors as width of the Gaussian distribution that we used to randomly extract the simulations parameters. Machine learning methods can not really assign an error to the estimate of a parameter so that we decided to quantify the uncertainty by analyzing the statistics of the test results. As shown in Fig.\,\ref{fig:lin_reg}, the constraints we thus get are comparable to those in \citetalias{planck2018} suggesting that high order statistics is as efficient as the CMB joint with other probes. However, in order to strengthen this promising result, it would be interesting to probe a much wider range in the parameter hyperspace to see whether the preference for multilinear regression we have found is genuine or an artifact of having used a so small range that all deviations from the fiducial can be parameterized through linear relations. It is entirely possible, indeed, that in this case more sophisticated machine learning methods would stand out as preferred ones, eventually improving the constraints.

A final point to discuss concerns the precision of the estimators measurements. As we have seen, higher scores ask for higher precision which can be obtained by averaging over a large number of maps. Ideally, one could generate as many maps as needed from the same initial simulation, but this is no more the case if one wants to rely on real maps. Indeed, for a \Euclid like survey, cutting the total 15000 sq deg survey area leads to $\sim 300$ maps which must be taken as an upper limit preventing to increase precision through averaging over an {\it ad libitum} number of maps. Luckily, Fig.\,\ref{fig:learning_curve} shows that, even keeping fixed the number of maps to $\sim 300$, we can still improve our predictions increasing the dataset size. For this project we were able to employ only $\sim \num{200000}$ hours of CPU time. The use of few million hours of computation time on a more powerful hardware (such as a national supercomputer) could allow us to build a dataset $10-100$ times larger. As an alternative, one could rely on a decent number of realistic N\,-\,body simulations to cut more maps investing the same amount of computational resources. Such a strategy, however, would ask for a preliminary narrowing of the parameter space so that additional probes should be used to avoid wasting time in exploring models already failing to fit other data.

Therefore, before this method can be applied to observations there are two main technical aspects that need to be addressed. First of all, the simulations must be realistic, including all the effects that we have been discussed above, and more specifically, they must be as close as possible to the actual data that we want to use in terms of survey characteristics such as source redshift distribution, noise, photo-z errors, and mask. The survey mask in particular, will determine the available area and the maximum number of maps that can be generated. Second, such simulations would require a significant amount of computation time so that the definition and the sampling of the cosmological parameter space must be optimized, e.g. with Latin hypercube sampling (\citealt{mckay1979}; \citealt{tang1993}; Euclid Collaboration \citeyear{euclid2019}).

\section{Conclusions}\label{sec:conclusions}

The search for new statistical methods able to shed further light on the nature and nurture of dark energy is becoming more and more important as the promise of unprecedented high quality data from Stage IV lensing surveys (such as \Euclid) moves towards reality. Motivated by this consideration, we have here investigated a machine learning approach to higher order statistics of the shear field. Going beyond second order better probes the non-Gaussianity imprinted on the shear field by the nonlinear collapse of structures hence allowing to alleviate degeneracy among cosmological parameters.

Our proposed method is innovative in three aspects. First, we directly work on the shear field as reconstructed from noisy galaxy ellipticity field which is the only quantity actually measured from images. This makes it possible to circumvent the non-trivial problem of map making, i.e, the need to reconstruct the convergence field from noisy shear data. As a second novelty, we have added some graph statistics measurements to the list of the estimators which (to the best of our knowledge) have never been used before in the context of WL studies, and on shear maps in particular. The third new aspect of the proposed methodology is the decision to use machine learning techniques to infer the almost complete set cosmological parameters solely from the shear higher order estimators. We have been motivated by the consideration that a theoretical formalism to compute the adopted quantities is available only for some of them and, in any case, based on a number of approximations and hypotheses that limits their application and may risk to introduce uncontrolled bias. The use of machine learning, on the contrary, is free of any assumption, flexible enough to include as many estimators as we want, and as reliable as the training set is. While the application of machine learning for the prediction of cosmological parameters in the context of WL is not a new concept {\it per se}, our work differs from the rest of the literature. In fact, previous WL machine learning studies have been limited to the use of convergence maps, to the variation of only two parameters ($\Omega_{\rm{M}}$,$\sigma_8$) in a wide range sampled in big steps, and to the training of neural networks. To our knowledge, this is the first work in which several machine learning methods were applied to noisy ellipticity maps, more than two cosmological parameters were made vary, and the contribution of each estimator to the measurements of each parameter was investigated.     

Using \texttt{CLASS} and \texttt{FLASK}, we produced $500$ simulated noisy ellipticity maps, at five redshift bins, for $1500$ sets of cosmological parameters, making each parameter vary randomly within $1\,\sigma$, and $2\,\sigma$ for a smaller sample, from the values measured by \citetalias{planck2018}. On each map we measured the HOM, MFs, Betti numbers, and graph statistics higher order estimators at different thresholds, for a total of $29$ features per redshift bin. We created several datasets to investigate how the size, the accuracy, and the smoothing of the training sample affects the results obtained on the test set. In the AVG100 dataset, we averaged each feature over $100$ maps (corresponding to a total area of $2500\,\rm{deg}^2$, obtaining five independent realizations for cosmology and a total of \num{7500} independent realizations) and applied Gaussian smoothing with $\theta_{s}=\{2\arcminute,4\arcminute,6\arcminute\}$. Both the AVG300 and AVG500 datasets contain one independent realizations for cosmology and a total of \num{1500} independent realizations, but in the first dataset we averaged each feature over $300$ maps (corresponding to a total area of $7500\,\rm{deg}^2$), with smoothing scales $\theta_{s}=\{2\arcminute,4\arcminute,6\arcminute\}$, while in the second the average was performed using $500$ maps (corresponding to a total area of $12500\,\rm{deg}^2$), for smoothing $\theta_{s}=2\arcminute$ only.

We performed the model selection comparing different machine learning algorithms and found out that the best performing model is also the simplest one, i.e., the linear regression. As we expected, the score decreases increasing the smoothing scale, and more severely so for the $\rm{w_0}$, $\sigma_8$, and $\rm{n_s}$ cosmological parameters, which appear to be more affected by the loss of information due to the smoothing, compared to $\Omega_{\rm{M}}$ and $\Omega_{\Lambda}$. We observed that the precision of the feature measurements (i.e. the signal-to-noise ratio) has to be favored over the number of independent realizations per cosmology in the training dataset because the score generally increases with the number of maps used to perform the averages, i.e. with the total survey area. In fact, we obtained a better performance with the AVG500 dataset, which contains only one realization per cosmology but a higher feature signal-to-noise ratio, compared to the AVG100 dataset, which contains five realizations per cosmology but lower feature signal-to-noise ratio.

We found the best scores for the AVG500 dataset with smoothing $\theta_{s}=2\arcminute$. We were able to accurately predict $\rm{w_0}$ and $\sigma_8$ with a score $\rm{R}^2 \sim 0.8$, followed by $\Omega_{\rm{M}}$ and $\Omega_{\Lambda}$ with $\rm{R}^2 \sim 0.7$, and $\rm{n_s}$ with $\rm{R}^2 \sim 0.3$. The remaining parameters, $\rm{H_0}$ and $\omega_{\rm{b}}$, could not be measured with our approach. On one hand this confirms the greatest sensitivity of WL to certain cosmological parameters, as expected from previous cosmic shear studies. On the other hand, considering the lack of external constraints on $\rm{H_0}$, $\omega_{\rm{b}}$, and $\rm{n_s}$, it could be surprising to an extent that one of the best measured parameters is indeed $\rm{w_0}$, even if this result could probably be due in part to the fact that we kept fixed $\rm{w_a}$, the parameter that controls the evolution of the dark energy equation of state. The other interesting aspect of this work consists in the investigation of the importance of each feature in the measurement of the different cosmological parameters. The new estimators that we introduced, the graph statistics, resulted to be very promising, contributing effectively to the prediction of all parameters (remarkably so for $\Omega_{\rm{M}}$ and $\Omega_{\Lambda}$), along with MFs that confirmed their utility even when applied to shear maps. The HOM are important for the measurement of $\rm{w_0}$ and especially of $\sigma_8$, while the Betti numbers contribute less compared to the other estimators. In terms of redshift, the majority of the information comes for low-medium $z$ for $\Omega_{\rm{M}}$ and $\Omega_{\Lambda}$, low $z$ for $\rm{n_s}$ and $\sigma_8$, and from all bins but with a peak at medium-high $z$ for $\rm{w_0}$.

We also discussed the limitations of this work, which consist mainly in the approximations made on the simulations side, the reduced redshift and cosmological parameters range used, and the limited dataset size that we were able to produce. This work was performed using only a couple hundred thousands of computation hours but with additional resources it would be possible to increase the size of the dataset and to explore a larger portion of the cosmological parameters hyperspace from which potentially more complex relations between the features and the labels could emerge. On the other hand, such resources could also be invested in the production of more realistic simulations. 

Finally, we want to stress the potential of this approach in terms of its flexibility. The fact that we do not need to develop a theoretical treatment of the statistics that we want to use, in order to express their relation to cosmology and their expected value for the universe we are considering, opens the way to explore different interesting estimators and even to create new ones. This method allows us to easily introduce new features and to study their relevance in the measurement of a specific cosmological parameter, along with the particular redshift range that we need to probe to access the majority of the information. Moreover, this regression model could be turned into a classification model that we could employ to distinguish between alternative cosmologies, such as $f(R)$ modified gravity models or really any model that differs from the standard $\Lambda$CDM model. We conclude that, considering the results that we obtained with this first and somehow rough attempt at the application of this method, we believe that it is worth to take further the investigation of this promising approach in the context of lensing higher order statistics analysis in future works. 

\section*{Acknowledgments}
CP and VFC are funded by Italian Space Agency (ASI) through contract
Euclid - IC (I/031/10/0)
and acknowledge financial contribution from the agreement
ASI/INAF/I/023/12/0. We acknowledge the support from the grant MIUR
PRIN 2015 Cosmology and Fundamental Physics: illuminating the Dark
Universe with Euclid.

\bibliographystyle{aa} 
\bibliography{ml} 

\begin{appendix}

\section{Models description}\label{app:models}

Here we briefly overview the methods that we used and compared in Section \ref{sec:model_sel}. We remind that we used the Scikit-learn \citep{sklearn} library implementation of all the listed algorithms. We refer to comprehensive machine learning books for the theoretical background \citep[e.g.][]{bishop2006, hastie2009, murphy2012} and to the python Scikit-learn library page on regression problems\footnote{\url{https://scikit-learn.org/stable/supervised_learning.html#supervised-learning}} for the technical description of each method. When possible, we will refer to more specific resources for the details on the particular algorithm implementation that we used, contained in the library.

\begin{itemize}
  \item Linear regression: linear model with as many coefficients as the number of features. It aims to minimize the residual sum of squares between the true labels and the labels predicted by the linear approximation. Calling $\mathbf{X}$ the feature matrix, $\mathbf{w}$ the coefficient vector, and $\mathbf{y}$ the labels, we want to solve a problem of the form
  \begin{equation*}
    \min_{w}{\lVert \mathbf{X}\,\mathbf{w} - \mathbf{y} \rVert}_{2}^{2}
  \end{equation*}
  \item Ridge regression: penalized linear model with as many coefficients as the number of features. It adds a L2-norm penalty term, which controls the size of the coefficients, to the residual sum of squares that has to be minimized. The penalty term is controlled by a hyperparameter $\alpha$. This corresponds to solving the problem
  \begin{equation*}
    \min_{w}{\lVert \mathbf{X}\,\mathbf{w} - \mathbf{y} \rVert}_{2}^{2} + \alpha {\lVert \mathbf{w} \rVert}_{2}^{2}
  \end{equation*}
  See \citet{rifkin2007} for theoretical and implementation details.
  \item Kernel Ridge regression: ridge regression with the application of a kernel, i.e. a function $k(\mathbf{x},\mathbf{x^\prime}):\mathcal{X} \times \mathcal{X} \rightarrow \mathbb{R}$ that measures similarity between any two points $\mathbf{x},\mathbf{x^\prime}$ of the feature space $\mathcal{X}$. It allows to learn a linear function in the space induced by the kernel which corresponds to a non-linear function in the original space. The model is determined by the kernel choice and by the regularization hyperparameter $\alpha$.
  \item Bayesian Ridge regression: probabilistic regression model with as many coefficients as the number of features. It imposes a prior over the coefficients $\mathbf{w}$ in the form of a spherical Gaussian
  \begin{equation*}
    p(\mathbf{w}|\lambda)=\mathcal{N}(\mathbf{w}|0,\lambda^{-1}I_{p})
  \end{equation*}
  and the output is assumed to be a Gaussian distribution around $\mathbf{X}\,\mathbf{w}$
  \begin{equation*}
    p(\mathbf{y}|\mathbf{X},\mathbf{w},\alpha)=\mathcal{N}(\mathbf{y}|\mathbf{X}\,\mathbf{w},\alpha)
  \end{equation*}
  where $\alpha$ and $\lambda$ are two regularization hyperparameters, which control the precision of the estimate and are computed from the data with the assumption of uninformative priors. The aim is to maximize the log marginal likelihood of the model. 
  \item Lasso regression: penalized linear model with sparse coefficients. It adds a L1-norm penalty term, which reduces the number of features used in the regression, to the residual sum of squares that has to be minimized. The penalty term is controlled by a hyperparameter $\alpha$. This corresponds to solving the problem
  \begin{equation*}
    \min_{w}{\lVert \mathbf{X}\,\mathbf{w} - \mathbf{y} \rVert}_{2}^{2} + \alpha {\lVert \mathbf{w} \rVert}_{1}
  \end{equation*}
  See \citet{tibshirani2010} and \citet{kim2008} for theoretical and implementation details.
  \item Support Vector Machine: Similar to Kernel Ridge regression but instead of the squared error function it uses the $\epsilon$-insensitive error function which learns a sparse model, ignoring errors which are smaller than $\epsilon$. This corresponds to solving the problem
  \begin{equation*}
    \min_{w}V_{\epsilon}( \mathbf{X}\,\mathbf{w} - \mathbf{y}) + \alpha {\lVert \mathbf{w} \rVert}_{2}^{2}
  \end{equation*}
  with $V_{\epsilon}$ the $\epsilon$-insensitive error function
  \begin{equation*}
  V_{\epsilon}(r)= \left \{ 
  \begin{array}{ll}
  \displaystyle{0}, & \displaystyle{|r|\leq \epsilon} \\
  & \\
  \displaystyle{|r|-\epsilon}, & \displaystyle{\text{otherwise}} \\
  \end{array}
  \right .
  \end{equation*}
  The model is therefore given by the choice of the kernel, the $\epsilon$ hyperparameter, and the regularization hyperparameter. See \citet{smola2004} for a detailed description of Support Vector Machine regression theory and algorithms.
  \item K Nearest Neighbors: the label corresponding to the set of input features is given by the mean of the label values of the $k$ nearest neighbors points in the feature space. The number of neighbors to consider is a hyperparameter of the model and different metrics can be used to calculate the distance between points in the feature space. We used the default standard Euclidean distance. See \citet{bentley1975} for the K-D Tree algorithm and \citet{omohundro1989} for the Ball Tree algorithm.  
  \item Gaussian Processes: given the features $\mathbf{x_i}$ and the corresponding label $y_i$ we want to find the function
  \begin{equation*}
    y_i = f(\mathbf{x_i})+\epsilon
  \end{equation*}
  where $\epsilon$ is the noise, assumed to be Gaussian $\mathcal{N}(0,\sigma_n)$. We need then to infer a distribution over functions given the data $p(f|\mathbf{X},\mathbf{y})$ and from that make predictions on test points $(\mathbf{x_*},y_*)$ calculating the mean of the conditional distribution $p(f_*|\mathbf{x_*},\mathbf{X},\mathbf{y})$. We can use Gaussian Processes to solve this problem. A Gaussian Process is a collection of random variables, any finite subset of which have a joint Gaussian distribution, and it is completely specified by its mean function and its covariance function. 
  In the noisy case, given a kernel, we can write the covariance of the prior distribution over the target functions $\mathbf{f}=f(\mathbf{x})$ as
  \begin{equation*}
    \rm{cov}(f(\mathbf{x_i}),f(\mathbf{x_j}))=k(\mathbf{x_i},\mathbf{x_j})+\sigma_n^2 \delta_{ij}
  \end{equation*} 
  where $k$ is a kernel function. In order to model the predictive distribution we can apply Bayes theorem and use the prior to condition the training data to model the joint distribution $p(\mathbf{f},\mathbf{f_*})$ of the training functions $\mathbf{f}$ and the functions in test points $\mathbf{f_*}$, which by definition of Gaussian Processes will be a joint Gaussian 
  \begin{equation*}
    \left(
    \begin{array}{l}
    \mathbf{f}\\
    \mathbf{f_*}\\
    \end{array}
    \right)
    = \mathcal{N} \sim \left (
    \begin{array}{ll}
    \mathbf{0}, & \left ( \begin{array}{ll} 
      K(\mathbf{X},\mathbf{X})+\sigma_n^2I & K(\mathbf{X},\mathbf{X_*}) \\
      K(\mathbf{X_*},\mathbf{X}) & K(\mathbf{X_*},\mathbf{X_*}) \\
    \end{array}
    \right )
    \end{array}
    \right )
  \end{equation*} 
  where $K$ is the covariance matrix and $K_{ij}=k(\mathbf{x_i},\mathbf{x_j})$. From $p(\mathbf{f},\mathbf{f_*})$ we can then calculate the posterior predictive distribution $p(f_*|\mathbf{x_*},\mathbf{X},\mathbf{y})$.
  The model is defined by the kernel choice. The kernel hyperparameters are fitted from the data. See \citet{rasmussen2006} for a theoretical overview and Algorithm 2.1 therein for the implementation details.
  \item Decision Tree: non-parametric method that creates a model applying a series of binary decision rules to the features, through as a series of \textit{nodes}. At each node, the sample is divided in two subsamples using the best split, i.e. the split that corresponds to the binary decision that minimizes the mean squared error between the true labels and the predicted labels, among all possible decisions (one for each feature). Each subsample is in turn split in two and the procedure is iterated until the terminal nodes, called \textit{leaf} nodes, are reached and a prediction is given. The structure created in this way is called \textit{tree}. The maximum depth of the tree is a hyperparameter of the model. See \citet{breiman1984} for a description of Classification and Regression Tree (CART) algorithms.   
  \item Random Forests: Ensemble of decision trees. Each tree is built from a bootstrap sample of the training data. The best split at each node of the tree is determined using a random subset of the input features. The prediction is given by averaging the results of all the trees in the ensemble. Among others, the number of trees, the maximum depth of each tree, and the number of features used to choose the split are hyperparameters of the model. See \citet{breiman2001} for a detailed description of Random Forests theory and implementation.  
  \item Gradient Boosting: Similar to Random Forests but at each iteration the current tree is trained using the residual error of the previous tree, in order to refine the prediction of the labels. The contribution of each tree is controlled by a learning rate hyperparameter. See \citet{fawcett2001} and \citet{friedman2002} for algorithm details.
\end{itemize}

\section{MSE and MAPE results}\label{app:mse_mape}

In Section\,\ref{sec:prediction}, we presented and discussed the results that we obtained performing the training and the prediction for each cosmological parameter on all three datasets, AVG100, AVG300, and AVG500 with smoothing $\theta_{s}=2\arcminute$ and on the AVG100 and the AVG300 datasets with the additional smoothing scales $\theta_{s}=\{4\arcminute,6\arcminute\}$, using the $\rm{R}^2$ score as the only metric to evaluate the different model performances. 

Here we expand on those results, focusing on the analysis of the $\rm{R}^2$ score values in conjunction with the mean squared error (MSE) and with the mean absolute percentage error (MAPE), which are defined as

\begin{equation}
\begin{aligned}
\rm{MSE} = & \frac{1}{\rm{N}}\sum{\left( y_{\rm{pred}}- y_{\rm{true}} \right)^2} \\
\rm{MAPE} = & \frac{1}{\rm{N}}\sum{\left | \frac{\left( y_{\rm{pred}}- y_{\rm{true}} \right)}{y_{\rm{true}}}\right |} \times 100
\end{aligned}
\end{equation}

where $y_{\rm{true}}$ are the true labels, $y_{\rm{pred}}$ are the predicted labels, and $\rm{N}$ is the total number of data points (i.e. independent realizations in the dataset).

In Table\,\ref{tab:app_lin_reg_smooth2},\ref{tab:app_lin_reg_smooth4}, and \ref{tab:app_lin_reg_smooth6}, we add to the scores already shown in Section\,\ref{sec:prediction} in Table\,\ref{tab:lin_reg_smooth2},\ref{tab:lin_reg_smooth4}, and \ref{tab:lin_reg_smooth6}, the MSE, and the MAPE that we obtained in each case. The dash in the $\rm{R}^2$ column represents a negative value of the score, we do not report the values of the MSE and MAPE in those cases. We notice that parameters that have a low-medium score ($\rm{R}^2<0.5$), as $\rm{H_0}$, $\rm{n_s}$, and $\sigma_8$ in some instances in Table\,\ref{tab:app_lin_reg_smooth2},\ref{tab:app_lin_reg_smooth4}, and \ref{tab:app_lin_reg_smooth6}, present a low value of the MSE and/or of the MAPE. This is somehow counter intuitive because we would expect high MSE and MAPE for low scores. 

We can explain this result rewriting Eq.\,\ref{eq:r2} using the following relation

\begin{equation}
\begin{aligned}
\rm{TSS} = & \rm{RSS} + \rm{ESS} \\
\rm{ESS} = & \sum{\left( y_{\rm{pred}}-\left< y_{\rm{true}} \right> \right)^2}
\end{aligned}
\end{equation}

where $\rm{RSS}$ is the residual sum of squares, $\rm{TSS}$ is the total sum of squares, as defined in Eq.\,\ref{eq:r2}, and $\rm{ESS}$ is the explained sum of squares or the sum of squares due to regression. Eq.\,\ref{eq:r2} then becomes

\begin{equation}
\rm{R}^2 = 1 - \frac{\rm{RSS}}{\rm{TSS}} = \frac{\rm{ESS}}{\rm{TSS}} \\
\end{equation}

which can be interpreted as the total variance explained by the model over the total variance, or the proportion of the variance in the true label values that is predictable from the features. In other words, the $\rm{ESS}$ and the $\rm{TSS}$ measure how much variation there is in the predicted label values and in the true label values, respectively.

This means that a low $\rm{R}^2$ score corresponds to $\rm{ESS} \ll \rm{TSS}$ so that the model will produce predicted values that are too close to the mean value of the true label distribution compared to the actual variance of the true label distribution. An example is a constant or almost constant model that always predicts the mean value of the true label distribution. However, if at the same time the true label distribution has a low $\rm{TSS}$ value, i.e. low variance (as is the case, considering the small range of variation of the cosmological parameters in the simulations), this model could lead to low MSE and MAPE, because on average the predicted labels will be close to the true labels, being both concentrated around the mean. This means that the MSE and MAPE alone have no real meaning in evaluating the performance of a model and have to be considered in conjunction with the $\rm{R}^2$ score.

We can look at Fig.\,\ref{fig:lin_reg} to better understand this concept. As explained in Section\,\ref{sec:prediction}, Fig.\,\ref{fig:lin_reg} shows the true labels versus the predicted labels, for each cosmological parameter, using the AVG100, the AVG300, and the AVG500 datasets, with smoothing $\theta_{s}=2\arcminute$. The dots represent individual predictions while the shaded areas correspond to the $1~\sigma$ region, obtained dividing the test sample into $10$ bins of the true label values and calculating the mean and standard deviation of the predictions inside each bin. The more the colored region for each given parameter aligns along the dashed black diagonal, the better will be the prediction obtained with such model. The black dots are the values of \citetalias{planck2018} parameters with the respective error, reported also on the $y-$axis as reference. We can see that $\rm{H_0}$ and $\omega_{\rm{b}}$ in all three versions, and $\rm{n_s}$ in the AVG100 version and slightly in the AVG300 version, fall in the situation described above. The predicted model, in fact, is represented by a more or less horizontal region and the individual predictions correspond to a cloud of dots concentrated around the mean value. This results in a bad model that is not able to predict the true label values over the entire range considered, especially for extreme values, which explains the low $\rm{R}^2$ score. Because the cloud of individual predictions though is centered around the mean with a small scatter, the MSE and MAPE have also a low value.

The remaining parameters show an almost perfect alignment along the diagonal, especially for the AVG500 version, confirming the good $\rm{R}^2$ score results presented in Table\,\ref{tab:app_lin_reg_smooth2}.

\begin{table*}
\begin{center}
\begin{tabular}{llllllllll}
\hline \hline
 & \multicolumn{3}{l}{AVG100, $\theta_{s}=2\arcminute$} & \multicolumn{3}{l}{AVG300, $\theta_{s}=2\arcminute$} & \multicolumn{3}{l}{AVG500, $\theta_{s}=2\arcminute$} \\
\hline
                  & $\rm{R}^2$  &  MSE         &  MAPE  &   $\rm{R}^2$        &  MSE   &  MAPE   &   $\rm{R}^2$        &  MSE   &  MAPE     \\
\hline
$\rm{H_0}$        & 0.16 & 0.30                 & 0.6\% & 0.09      & 0.32                 & 0.6\% & 0.03      & 0.34                 & 0.7\% \\
$\omega_{\rm{b}}$ & -    & -                    & -     & -         & -                    & -     & -         & -                    & -     \\
$\Omega_{\rm{M}}$ & 0.61 & $2.30\times 10^{-5}$ & 1.2\% & 0.64      & $2.29\times 10^{-5}$ & 1.2\% & 0.70      & $1.89\times 10^{-5}$ & 1.1\% \\
$\Omega_\Lambda$  & 0.61 & $2.30\times 10^{-5}$ & 0.5\% & 0.64      & $2.29\times 10^{-5}$ & 0.5\% & 0.70      & $1.89\times 10^{-5}$ & 0.5\% \\
$\rm{w_0}$        & 0.65 & $7.34\times 10^{-4}$ & 2.0\% & 0.75      & $4.57\times 10^{-4}$ & 1.6\% & 0.81      & $3.51\times 10^{-4}$ & 1.5\% \\
$\rm{n_s}$        & 0.16 & $2.22\times 10^{-5}$ & 0.4\% & 0.23      & $1.78\times 10^{-5}$ & 0.3\% & 0.32      & $1.58\times 10^{-5}$ & 0.3\% \\
$\sigma_8$        & 0.56 & $2.74\times 10^{-5}$ & 0.5\% & 0.73      & $1.68\times 10^{-5}$ & 0.4\% & 0.80      & $1.25\times 10^{-5}$ & 0.4\% \\  
\hline \hline 
\end{tabular}
\end{center}
\caption{$\rm{R}^2$ score, MSE, and MAPE for the three datasets, AVG100, AVG300, and AVG500 with smoothing scale $\theta_{s}=2\arcminute$, for each cosmological parameter. The dash in the $\rm{R}^2$ column represents a negative value of the score, we do not report the values of the MSE and MAPE in those cases. As expected, increasing the number of maps per simulation improves the performance of the algorithm.}
\label{tab:app_lin_reg_smooth2}

\begin{center}
\begin{tabular}{lllllll}
\hline \hline
 & \multicolumn{3}{l}{AVG100, $\theta_{s}=4\arcminute$} & \multicolumn{3}{l}{AVG300, $\theta_{s}=4\arcminute$} \\
\hline
                  & $\rm{R}^2$  &  MSE         &  MAPE  &   $\rm{R}^2$        &  MSE   &  MAPE   \\
\hline
$\rm{H_0}$        & 0.16 & 0.35                 & 0.7\% & -         & -                    & -     \\
$\omega_{\rm{b}}$ & -    & -                    & -     & -         & -                    & -     \\
$\Omega_{\rm{M}}$ & 0.55 & $2.57\times 10^{-5}$ & 1.2\% & 0.60      & $2.51\times 10^{-5}$ & 1.2\% \\
$\Omega_\Lambda$  & 0.55 & $2.57\times 10^{-5}$ & 0.6\% & 0.60      & $2.51\times 10^{-5}$ & 0.5\% \\
$\rm{w_0}$        & 0.62 & $8.53\times 10^{-4}$ & 2.3\% & 0.68      & $5.82\times 10^{-4}$ & 1.9\% \\
$\rm{n_s}$        & 0.06 & $2.60\times 10^{-5}$ & 0.4\% & -         & -                    & -     \\
$\sigma_8$        & 0.52 & $3.17\times 10^{-5}$ & 0.5\% & 0.64      & $2.27\times 10^{-5}$ & 0.5\% \\  
\hline \hline 
\end{tabular}
\end{center}
\caption{Same as in Table\ref{tab:lin_reg_smooth2} but for the AVG100 and AVG300 datasets with smoothing scale $\theta_{s}=4\arcminute$.}
\label{tab:app_lin_reg_smooth4}

\begin{center}
\begin{tabular}{lllllll}
\hline \hline
 & \multicolumn{3}{l}{AVG100, $\theta_{s}=6\arcminute$} & \multicolumn{3}{l}{AVG300, $\theta_{s}=6\arcminute$} \\
\hline
                  & $\rm{R}^2$  &  MSE         &  MAPE  &   $\rm{R}^2$        &  MSE   &  MAPE   \\
\hline
$\rm{H_0}$        & 0.18 & 0.34                 & 0.7\% & 0.06      & 0.33                 & 0.7\% \\
$\omega_{\rm{b}}$ & -    & -                    & -     & -         & -                    & -     \\
$\Omega_{\rm{M}}$ & 0.55 & $2.57\times 10^{-5}$ & 1.3\% & 0.61      & $2.46\times 10^{-5}$ & 1.2\% \\
$\Omega_\Lambda$  & 0.55 & $2.57\times 10^{-5}$ & 0.6\% & 0.61      & $2.46\times 10^{-5}$ & 0.5\% \\
$\rm{w_0}$        & 0.60 & $9.17\times 10^{-4}$ & 2.3\% & 0.60      & $7.37\times 10^{-4}$ & 2.1\% \\
$\rm{n_s}$        & 0.04 & $2.68\times 10^{-5}$ & 0.4\% & -         & -                    & -     \\
$\sigma_8$        & 0.47 & $3.50\times 10^{-5}$ & 0.6\% & 0.52      & $3.00\times 10^{-5}$ & 0.5\% \\  
\hline \hline 
\end{tabular}
\end{center}
\caption{Same as in Table\ref{tab:lin_reg_smooth2} but for the AVG100 and AVG300 datasets with smoothing scale $\theta_{s}=6\arcminute$. Comparing this results with those in Table\,\ref{tab:lin_reg_smooth2} and Table\,\ref{tab:lin_reg_smooth4}, we observe that overall the score decreases with increasing smoothing scale.}
\label{tab:app_lin_reg_smooth6}
\end{table*}

\end{appendix}

\end{document}